\documentclass[11pt]{article}

\usepackage{authblk}
\usepackage{listings}
\usepackage[numbers]{natbib}
\usepackage{hyperref}
\usepackage{xcolor}

\definecolor{hrefcolor}{rgb}{0.0,0.0,0.8}
\newcommand{\linkcolor}{hrefcolor}

\hypersetup{pdfstartview=FitR}
\hypersetup{pdftitle={Python Framework for HP Adaptive Discontinuous Galerkin Method for Two Phase Flow in Porous Media},
            pdfauthor={A. Dedner, B. Kane, R. Kloefkorn, M. Nolte},
            linkbordercolor=\linkcolor,
            urlbordercolor=\linkcolor,
            citebordercolor=\linkcolor,
            runbordercolor=\linkcolor,
            menubordercolor=\linkcolor,
            filebordercolor=\linkcolor,
            runcolor=\linkcolor
            baseurl={http://dune-project.org}
}
\hypersetup{pdfborderstyle={/S/U/W 1}}

\newcommand{\journal}[1]{}
\newcommand{\corref}[1]{}
\newcommand{\cortext}[2][]{}
\newcommand{\ead}[2][]{}%\texttt{#1}}
\newcommand{\address}[2][]{\affil[#1]{#2}}
\newcommand{\sep}{, }
\newcommand{\putmaketitle}{\maketitle}
\newenvironment{frontmatter}{
}
{
}
\newenvironment{keyword}{\textbf{Keywords:}
}{}
\newcommand{\refsection}{}
\date{April 15, 2018}
\usepackage{amssymb}
\usepackage{amsthm}

\usepackage{lineno}

\usepackage{comment}

\usepackage[utf8]{inputenc}
\usepackage{tikz}
\usetikzlibrary{shapes,snakes}
\usepackage{commath}
\usepackage{pgfplots}
\usetikzlibrary{patterns}

\usepackage{afterpage}
\usepackage{a4wide}
\usepackage{amsmath}
\usepackage{amsfonts}
\usepackage{epsfig}
\usepackage{graphics}
\usepackage{stmaryrd}
\usepackage{algorithm,algpseudocode, algorithmicx}
\usepackage{python}
\usepackage{xcolor}
\usepackage{lipsum}                     % Dummytext
\usepackage{xargs}                      % Use more than one optional parameter in a new commands
\usepackage[colorinlistoftodos,prependcaption,textsize=tiny]{todonotes}
\newcommandx{\unsure}[2][1=]{\todo[linecolor=red,backgroundcolor=red!25,bordercolor=red,#1]{#2}}
\newcommandx{\change}[2][1=]{\todo[linecolor=blue,backgroundcolor=blue!25,bordercolor=blue,#1]{#2}}
\newcommandx{\info}[2][1=]{\todo[linecolor=OliveGreen,backgroundcolor=OliveGreen!25,bordercolor=OliveGreen,#1]{#2}}
\newcommandx{\improvement}[2][1=]{\todo[linecolor=Plum,backgroundcolor=Plum!25,bordercolor=Plum,#1]{#2}}
\newcommandx{\thiswillnotshow}[2][1=]{\todo[disable,#1]{#2}}

\newcommand{\wavg}[1]{\big\{ #1 \big\}_\omega}
\newcommand{\jump}[1]{\llbracket  #1 \rrbracket}
\newcommand{\bars}{\bar{s}}

\journal{Applied Mathematical Modelling}

\makeatletter
\algrenewcommand\ALG@beginalgorithmic{\scriptsize}
\makeatother

\begin{document}

\begin{frontmatter}


\title{Python Framework for HP Adaptive Discontinuous Galerkin Method for Two Phase Flow in Porous Media}
\author[a]{Andreas Dedner\corref{cor1}}
\ead{a.s.dedner@warwick.ac.uk}
\ead[url]{https://warwick.ac.uk/fac/sci/maths/people/staff/andreas\_dedner/}
\address[a]{University of Warwick, Coventry CV4 7AL UK}
%\address[a]{University of Warwick, Coventry CV4 7AL UK}
\cortext[cor1]{Corresponding author}
\author[b]{Birane Kane}
\ead{birane.kane@ians.uni-stuttgart.de}
\ead[url]{http://www.ians.uni-stuttgart.de/nmh/kane}
\address[b]{University of Stuttgart, Germany}
\author[c]{Robert Kl\"ofkorn}
\ead{robert.kloefkorn@iris.no}
\ead[url]{www.iris.no}
\address[c]{International Research Institute of Stavanger, Norway}
\author[d]{Martin Nolte}
\ead{nolte.mrtn@gmail.com}
\ead[url]{https://aam.uni-freiburg.de/mitarb/ehemalige/nolte/index.html}
\address[d]{University of Freiburg, Germany}

\putmaketitle
%\date{April 15, 2018}
%\maketitle 

\begin{abstract}
  In this paper we present a framework for solving two phase flow problems in
  porous media. The discretization is based on a Discontinuous Galerkin
  method and includes local grid adaptivity and local choice of polynomial
  degree. The method is implemented using the new Python frontend
  \texttt{Dune-FemPy} to the open source framework Dune. The code used for the
  simulations is made available as Jupyter notebook and can be used through
  a Docker container. We present a number of time stepping approaches
  ranging from a classical IMPES method to fully coupled implicit scheme.
  The implementation of the discretization is very flexible allowing for
  test different formulations of the two phase flow model and adaptation
  strategies.
\end{abstract}

\begin{keyword}
DG\sep hp-adaptivity\sep Two-phase flow\sep IMPES\sep Fully implicit\sep 
Dune\sep Python\sep Porous media



\end{keyword}

\end{frontmatter}

% \linenumbers

%\section{Introduction}
%\label{sec:introduction}

\section{Introduction}
\label{sec:introduction}
Simulation of  multi-phase flows and transport processes in porous media requires careful numerical treatment due to the strong heterogeneity of the underlying porous medium.
The spatial discretization requires locally conservative methods in order to be able to follow small concentrations \cite{bastian1999numerical}.
Discontinuous Galerkin (DG) methods, Finite Volume methods and Mixed Finite Element methods are examples of discretization techniques achieving 
local conservation at the element level \cite{di2011mathematical}.
Application of DG methods to incompressible two-phase flow started within the framework provided by a decoupled approach called Implicit Pressure Explicit Saturation (IMPES) where first a pressure equation is solved implicitly and then the saturation is advanced by an explicit time stepping scheme.
 Upwinding, slope limiting techniques, and sometimes $\mathbb{H}$(div)-projection  were required in order to remove unphysical oscillations and to ensure convergence to a solution. 

In the fully implicit and fully coupled approach, the mass balances are usually discretized in time by the implicit Euler method, resulting in a fully coupled system of nonlinear equations that has to be solved at each time step.
The main advantage of a fully implicit scheme is the possibility of using significantly larger time step sizes, which can be crucial in view of long-term scenarios like atomic waste disposal.
Commonly, rather simple, yet robust space-discretization schemes, like cell-centered or vertex-centered finite volume schemes, 
are used \cite{bastian1999numerical,bastian1999efficient}.
Fully implicit DG schemes have been proposed in \cite{eps:07} and \cite{eps:09}, where the schemes are formulated in two space dimensions for incompressible fluid phases and numerical tests are performed without any kind of adaptivity.

Bastian introduced in \cite{bastian2014fully} a fully coupled symmetric interior penalty DG scheme for incompressible two-phase flow based on a  wetting-phase potential and capillary potential formulation.
Discontinuity in capillary-pressure functions is taken into account by incorporating the interface conditions into the penalty terms for the capillary potential.
Heterogeneity in absolute or intrinsic permeability is treated by weighted averages.
A higher-order diagonally implicit Runge-Kutta method in time is used and there is neither post processing of the velocity nor slope limiting.
Only piecewise linear and piecewise quadratic functions are employed and no adaptive method is considered.

A general abstract framework allowing for an a-posteriori estimator for porous-media two-phase flow problem was introduced by Vohralik et al.~\cite{vohralik2013posteriori}.
This paved the way for an h-adaptive strategy for homogeneous two-phase flow problems~\cite{cances2014posteriori}.
However, it has not been applied to DG methods so far.

Finally, Darmofal et al. introduced recently a space-time discontinuous Galerkin h-adaptive framework for 2d reservoir flows.
Implicit estimators are derived through the use of dual problems~\cite{becker1996feed,becker2001optimal} and a higher-order discretization is performed on anisotropic, unstructured meshes.
Unfortunately, application to 3d problems and hp-adaptive strategies haven't been considered yet.

In this paper, we implement and evaluate numerically interior penalty DG methods for incompressible, immiscible, two-phase flow.
We consider strongly heterogeneous porous media, anisotropic permeability tensors and discontinuous capillary-pressure functions.
We write the system in terms of a phase-pressure/phase-saturation formulation.
%TODO Aren't we only using Backward Euler here
%A backward Euler scheme in time is combined with an Interior Penalty DG discretization in space.

Adams-Moulton schemes of first or second order in time are combined with an Interior Penalty DG discretization in space.
This implicit space time discretization leads to a fully coupled nonlinear system requiring to build a Jacobian matrix at each time step for the Newton-Raphson method. 
%We also implement an  implicit iterative scheme for the numerical simulation of two-phase flow in porous media was presented in \cite{anna_thesisk}. The scheme is based on the iterative IMPES approach and treats the capillary pressure term implicitly to ensure stability. This is accomplished by thelinear approximation for the capillary pressure gradient which involves the saturation function on both current and previous iterations.  

 %We also implement iterative IMPES schemes and implicit iterative scheme for the numerical simulation of two-phase flow in porous media. The  implicit iterative scheme is based on the iterative IMPES approach \cite{} and treats the capillary pressure term implicitly to ensure stability. 
%This is accomplished by the linear approximation for the capillary pressure gradient which involves the saturation function on both current and previous iterations.  
%We include in our implementation local mesh adaptivity on non-conforming grids. 
 %
This paper extends our previous work in \cite{kanedune,kaneetal} and \cite{kane2017hp}.
We consider here new hp-adaptive strategies and compare the fully implicit scheme with the iterative IMPES scheme and the implicit iterative scheme.
The implicit iterative scheme is based on the iterative IMPES approach presented in \cite{kvashchuk:15} and treats the capillary pressure term implicitly to ensure stability.
We also provide a more comprehensive model framework allowing to conveniently implement and compare various two-phase flow formulations. 

The implementation is based on the open-source PDE software framework \texttt{Dune-FemPy}, which is a Python frontend for \texttt{Dune-Fem}~\cite{dunefempaper} 
based on the new \texttt{Dune-Python} module
\cite{dednernolte2018dnepython} and which adds support for the Unified Form Language~\cite{alnaes:14}. 
It allows for a compact, legible presentation of the different discretizations under consideration.
We combine \texttt{Dune-FemPy} with Jupyter~\cite{kluyver:16} and Docker~\cite{boettinger:15} to ensure reproducibility of our numerical experiments.
The adaptive grid implementation is based on \texttt{Dune-Alugrid}~\cite{alkamper2016dune} and parts of the 
stabilization mechanisms used are provided by
\texttt{Dune-Fem-DG}~\cite{dunefemdg:17}.

The rest of this document is organised as follows.
In Section~\ref{sec:model}, we describe the two-phase flow model.
The DG discretization is introduced in Section~\ref{sec:discretization}.
Numerical examples are provided in Section~\ref{sec:results}.
Conclusions are drawn in the last section.

%\section{Mathematical Model}
%\label{sec:model}

% here we should describe the mathematical model (A,B) resulting in the
% form -div(D_{pp}(p,s)grad(p) + ....), i.e., the form used for the
% dune-fem-howto
% I moved the p_c function to the problem class so the model only uses P.p_c
% One could first write down the model with the given problem parameters
% (K,p_c,etc), explain the quantities (s_n,p_w etc). In a second part then
% descript the 'problem' class perhaps even here providing the details of
% the AnisotropicLens problem?

\section{Mathematical Model}
\label{sec:model}
This section introduces the mathematical formulation of a two-phase porous-media flow. In all that follows, we assume that the flow is immiscible and incompressible with no mass transfer between phases.

\subsection{Two-phase flow formulation}
Let $\Omega$ be a polygonal bounded domain in $\mathbb{R}^d$, $d\in \{2,3\}$, with Lipschitz boundary $\partial \Omega$ and let $T \in \mathbb{R}_+$. The flow of the wetting phase (e.g. water) and the nonwetting phase (e.g. oil, gas) is described by Darcy's law and the continuity equation (e.g. balance of mass) for each phase $\alpha \in \{ w,n\}$\cite{helmig1997multiphase}. In all that follows, we denote with subscript $w$ the wetting phase and with subscript $n$ the nonwetting phase. 
The unknown variables are the phase pressures $p_w,\ p_n: \Omega \times (0, T)  \rightarrow \mathbb{R}$  and the phase saturations $s_w,\ s_n: \Omega \times (0, T)  \rightarrow \mathbb{R}$. 
 For each phase $\alpha \in \{ w,n\}$, the Darcy velocity $\mathbf{v}_{\alpha}: \Omega \times (0, T)  \rightarrow \mathbb{R}^d$ is given by
\begin{align}\label{eq:Darcy}
\mathbf{v}_{\alpha} = - \lambda_{\alpha} \mathbb{K} (\nabla p_{\alpha} - \rho_{\alpha} \mathbf{g}) \quad \quad  \mbox{in} \ \Omega \times (0,T)
\end{align}
where  $\lambda_{\alpha}: \Omega \times (0, T)  \rightarrow \mathbb{R}$ is the phase  mobility, $\mathbb{K}: \Omega  \rightarrow \mathbb{R}^{d\times d}$ is the absolute or intrinsic permeability tensor of the porous medium, $\rho_{\alpha}: \Omega \times (0, T)  \rightarrow \mathbb{R}$ is the phase density, and $\mathbf{g} \in \mathbb{R}^d$ is the constant gravitational vector. \\
Phase mobilities $\lambda_{\alpha}: \Omega \times (0, T)  \rightarrow \mathbb{R}$ are defined by 
 \begin{align}\label{Mobility} 
 \lambda_{\alpha}=\frac{ k_{r\alpha}}{\mu_{\alpha}} \mbox{, } \quad \alpha \in\{ w,n\},
\end{align}
where $\mu_{\alpha}$ is the constant phase viscosity and $k_{r\alpha}: \Omega \times (0, T)  \rightarrow \mathbb{R}$ is the relative permeability of phase $\alpha$. The relative permeabilities are functions that depend nonlinearly on the phase saturation (i.e.  $k_{r\alpha}=k_{r\alpha}(s_{\alpha})$). Models for the relative permeability are the van-Genuchten model~\cite{van1980closed} and the Brooks-Corey model~\cite{brooks1964hydraulic}.
 For example, in the Brooks-Corey model,
\begin{eqnarray}
k_{rw}(s_{n,e})= (1-s_{n,e})^{\frac{2+3\theta}{\theta}} \mbox{,}    \hspace{4mm}k_{rn}(s_{n,e})= (s_{n,e})^2 (1-(1-s_{n,e})^{\frac{2+\theta}{\theta}}),
\end{eqnarray}
where  the effective saturation $s_{\alpha, e}$ is
\begin{eqnarray}
s_{\alpha, e}=\frac{s_{\alpha}-s_{\alpha,r}}{1-s_{w,r}-s_{n,r}}, \hspace{4mm}\forall \alpha \in \{w,n \}. 
\end{eqnarray}
%
%\begin{eqnarray}
%s_{\alpha, e}=\cutoff{ \frac{s_{\alpha}-s_{\alpha,r}}{1-s_{w,r}-s_{n,r}} }.
%\end{eqnarray}
%
Here, $s_{\alpha,r}$, $\alpha \in \{w,n\}$, are the phase residual saturations. The parameter $\theta \in [0.2, 3.0]$ is a result of the inhomogeneity of the medium.\\
%We define the total velocity $\mathbf{v}_t: \Omega \times (0, T)  \rightarrow \mathbb{R}^d$ and the total mobility $\lambda_t: \Omega \times (0, T)  \rightarrow \mathbb{R}$ as
%%
%\begin{align}
%  \mathbf{v}_t&=\sum\limits_{\alpha \in \{ w,n\}}\mathbf{v}_{\alpha}\label{totalvelocity},\\
%  \lambda_t&=\sum\limits_{\alpha \in \{ w,n\}}\lambda_{\alpha}\label{totalmobility}.
%\end{align}
%
For each phase $\alpha \in\{ w,n\}$, the balance of mass yields the saturation equation
\begin{align}\label{eq:contineq}
 \phi \frac{\partial (\rho_\alpha s_{\alpha})}{\partial t} + \nabla \cdot (\rho_\alpha \mathbf{v}_{\alpha}) & =\rho_\alpha q_{\alpha},
\end{align}
 where $\phi: \Omega  \rightarrow \mathbb{R}$ is the porosity, $q_{\alpha}: \Omega \times (0, T)  \rightarrow \mathbb{R}$ is a source or sink term (e.g. wells located inside the domain in the case of a reservoir problem).\\
In addition to \eqref{eq:Darcy} and \eqref{eq:contineq} the following closure relations must also be satisfied:
\begin{align}
s_w+s_n&=1,\label{eq:sumsat} \\
p_{n} - p_{w}& = p_c(s_n),\label{eq:CapillaryPressure}
\end{align}
%where $p_c = p_c(s_w)$
where $ p_c(s_n): \Omega \times (0, T)  \rightarrow \mathbb{R}$ is the capillary pressure, a function of the phase saturation. For the Brooks-Corey formulation,
\begin{eqnarray}
p_c(s_{n,e}) = p_d (1-s_{n,e})^{-1/\theta}.\label{BCcappress}
\end{eqnarray}
Here, $p_d \ge 0$ is the constant entry pressure, needed to displace the fluid from the largest
pore.
In summary, the immiscible, incompressible two-phase flow formulation is
\begin{align}\label{eq:twophasesyst1}
\mathbf{v}_{\alpha} = - \lambda_{\alpha} \mathbb{K} (\nabla p_{\alpha}&-\rho_{\alpha} \mathbf{g}) \mbox{, } \quad \alpha \in\{ w,n\},\\ 
 \phi  \frac{\partial s_{\alpha}}{\partial t} + \nabla \cdot ( \mathbf{v}_{\alpha})  &= q_{\alpha}  \mbox{,} \quad \alpha \in\{ w,n\},\label{eq:twophasesyst2}\\
s_w+s_n &=1 ,\label{eq:twophasesyst3} \\
p_{n} - p_{w} &= p_{c}, \label{eq:twophasesyst4}
\end{align}
where we search for the phase pressures $p_\alpha$ and the phase saturations $s_\alpha$, $\alpha \in\{ w,n\}$.

%\subsection{Wetting-phase-pressure / nonwetting-phase-saturation formulation}
\subsection{Model A: Wetting-phase-pressure/nonwetting-phase-saturation formulation}

Considering the phases are incompressible (i.e. the densities $\rho_\alpha$ are constant), we get a total fluid conservation equation by summing the two mass balance equations from \eqref{eq:twophasesyst2}, 
\begin{align*}
\phi \frac{\partial ( s_{n} + s_{w})}{\partial t} + \nabla \cdot ( \mathbf{v}_{n} +  \mathbf{v}_{w}) & = q_{n} + q_{w}.
\end{align*}
Thanks to relation \eqref{eq:twophasesyst3},
\begin{align*}
\nabla \cdot ( \mathbf{v}_{n} +  \mathbf{v}_{w}) & = q_{n} + q_{w}.
\end{align*}
From relation \eqref{eq:twophasesyst1} we have
\begin{align*}
- \nabla \cdot (  \lambda_{n} \mathbb{K} (\nabla p_{n}-\rho_{n} \mathbf{g}) + \lambda_{w} \mathbb{K} (\nabla p_{w}-\rho_{w} \mathbf{g})) &= q_{n} + q_{w}.
\end{align*}
The last closure relation \eqref{eq:twophasesyst4} allows to write
\begin{align*}
- \nabla \cdot (  \lambda_{n} \mathbb{K} (\nabla p_{c}+ \nabla p_{w} -\rho_{n} \mathbf{g}) + \lambda_{w} \mathbb{K} (\nabla p_{w}-\rho_{w} \mathbf{g})) &= q_{n} + q_{w}.
\end{align*}
Finally, 
\begin{align*}
 -\nabla\cdot \biggl(( \lambda_w +  \lambda_n) \mathbb{K} \nabla p_{w} +  \lambda_n \mathbb{K} \nabla p_{c} - (\rho_w \lambda_w + \rho_n \lambda_n) \mathbb{K} \mathbf{g}\biggr)&= q_w + q_n.
\end{align*}
To complete our system, we consider as second equation the nonwetting phase conservation relation
\begin{align*}
\phi \frac{\partial s_{n} }{\partial t} + \nabla \cdot \mathbf{v}_{n}  & = q_{n}.
\end{align*}
Using relation \eqref{eq:twophasesyst1} and \eqref{eq:twophasesyst4} yields
\begin{align*}
 \phi \frac{\partial  s_n}{\partial t} - \nabla \cdot \biggl( \lambda_n \mathbb{K} (\nabla p_{w}-\rho_n \mathbf{g})\biggr) - \nabla \cdot \biggl( \lambda_n \mathbb{K} \nabla p_{c}\biggr)  &=  q_{n}.
 \end{align*}
 We get therefore a system of two equations with two unknowns $p_w$ and $s_n$,
%Equations \eqref{twophasesyst1} -\eqref{twophasesyst4} can be formulated as
\begin{align}\label{eq:1presswetnonwetsat}
 -\nabla\cdot \biggl( (\lambda_w + \lambda_n) \mathbb{K} \nabla p_{w} +  \lambda_n \mathbb{K} \nabla p_{c} - (\rho_w \lambda_w + \rho_n \lambda_n) \mathbb{K} \mathbf{g}\biggr)&= q_w + q_n,\\
 \phi \frac{\partial  s_n}{\partial t} - \nabla \cdot \biggl( \lambda_n \mathbb{K} (\nabla p_{w}-\rho_n \mathbf{g})\biggr) - \nabla \cdot \biggl( \lambda_n \mathbb{K} \nabla p_{c}\biggr)  &=  q_{n}.\label{eq:2presswetnonwetsat}
\end{align}
%
%From the constitutive relations \eqref{eq:close1} and \eqref{eq:close2}, we can rewrite the two-phase flow problem as a system of two equations  with two unknowns $p_w$ and $s_n$,
%\begin{eqnarray}\label{}
%\begin{aligned}\label{syst:presswetnonwetsatini}
% -\nabla\cdot \biggl( \lambda_t  \mathbb{K} \nabla p_{w} + \lambda_n \mathbb{K} \nabla p_{c} - (\rho_w\lambda_w + \rho_n \lambda_n) \mathbb{K} \mathbf{g} \biggr)&= q_w + q_n,\\
% \phi \frac{\partial  s_n}{\partial t} - \nabla \cdot \biggl( \lambda_n \mathbb{K} (\nabla p_{w}-\rho_n \mathbf{g})\biggr) - \nabla \cdot \biggl( \lambda_n \mathbb{K} \nabla p_{c} \biggr)  &=q_{n}.
%\end{aligned}
%\end{eqnarray}
%Here, $\lambda_t = \lambda_w + \lambda_n$ denotes the total mobility. 

Substituting $\nabla p_{c}=p_c^{\prime}(s_n) \nabla s_n$ for $\nabla p_{c}$ as in \cite{helmig1997multiphase, helmig1998comparison}, the system \eqref{eq:1presswetnonwetsat}-\eqref{eq:2presswetnonwetsat} becomes
%\begin{eqnarray}\label{}
\begin{align}\label{syst:1presswetnonwetsat}
 -\nabla\cdot \biggl( (\lambda_w + \lambda_n)  \mathbb{K} \nabla p_{w} + \lambda_n p'_c  \mathbb{K} \nabla s_{n} - (\rho_w\lambda_w + \rho_n \lambda_n) \mathbb{K} \mathbf{g}\biggr)&= q_w + q_n,\\
 \phi \frac{\partial  s_n}{\partial t} - \nabla \cdot \biggl( \lambda_n \mathbb{K} (\nabla p_{w}-\rho_n \mathbf{g})\biggr) - \nabla \cdot \biggl( \lambda_n p'_c \mathbb{K} \nabla s_{n}\biggr)  &=q_{n}.\label{syst:2presswetnonwetsat}
\end{align}
%\end{eqnarray}
%where $\lambda_c(s_n)= \lambda_n(s_n) p_c^{\prime}(s_n)$   is such that $\lambda_c(s_n) <0$.

In order to have a complete system, we add appropriate boundary and initial conditions. Thus, we assume that the boundary of the system is divided into disjoint sets  such that $\partial \Omega =  \Gamma_{ D} \cup  \Gamma^{}_{N}$.  We denote by $\nu$ the outward normal to $\partial \Omega$ and set
%
%\begin{align}\label{bndcond}
%s_n(\cdot,0)&=s^{0}_n(\cdot)   \mbox{,} \qquad  p_{w}(\cdot,0)=p^{0}_{w}(\cdot) \quad  \qquad \qquad     \mbox{in} \ \Omega, \\
% p_w&=p_{w,D} \mbox{,} \qquad  \qquad \quad s_n=s_{ D} \qquad    \qquad  \qquad    \mbox{on} \ \Gamma^{D}_{} \times (0, T), \\
%\mathbf{v}_\alpha \cdot \nu &= J_\alpha \mbox{,} \qquad \qquad \quad \ J_t= \sum_{\alpha \in \{w,n\}} J_\alpha \qquad \quad  \mbox{on}  \ \Gamma^{N}_{} \times (0, T).
%\end{align}
%
\begin{alignat*}{3}
   p_{w}(\cdot,0)&=p^{0}_{w}(\cdot)  ~,\qquad &
s_n(\cdot,0)&=s^{0}_n(\cdot) ~,\qquad & \mbox{in}& \ \Omega,\\
 p_w&=p_{w,D}  ~,\qquad &
 s_n&=s_{ D} ~,\qquad & \mbox{on}& \ \Gamma^{D}_{} \times (0, T),\\
\mathbf{v}_\alpha \cdot \nu &= J_\alpha  ~,\qquad &
 J_t&= \sum_{\alpha \in \{w,n\}} J_\alpha ~,\qquad & \mbox{on}&  \ \Gamma^{N}_{} \times (0, T).
\end{alignat*}
Here, $J_\alpha \in \mathbb{R}$, $\alpha \in \{w,n\}$, is the inflow,  $s^{0}_n, \  p^{0}_w, \ s_{ D}$, and $\ p_{ w,D}$ are real numbers.
In order to make $p_w$ uniquely determined, we require $\Gamma_{ D}\neq \emptyset$.
%% here we should describe the mathematical model (A,B) resulting in the
%% form -div(D_{pp}(p,s)grad(p) + ....), i.e., the form used for the
%% dune-fem-howto
%% I moved the p_c function to the problem class so the model only uses P.p_c
%% One could first write down the model with the given problem parameters
%% (K,p_c,etc), explain the quantities (s_n,p_w etc). In a second part then
%% descript the 'problem' class perhaps even here providing the details of
%% the AnisotropicLens problem?
\subsection{General model framework}
We provide here a unified model framework allowing for the representation of the models introduced in the previous sections,
% Assume model of the form
\begin{align} \label{eq:1generalFramework}
  -\nabla \cdot \biggl(  A_{pp}(s)\nabla p + A_{ps}(s)\nabla s + G_p(s) \biggr) = q_p, \\
  \Phi \partial_t s
  -\nabla \cdot \biggl( A_{sp}(s)(\nabla p-P_g) + A_{ss}(s)\nabla s + G_s(s) \biggr) = q_s.\label{eq:2generalFramework}
\end{align}
The model is described once the physical parameter functions $A$, $G$, and $P_g$ are known.
%The method is completely described once the physical parameter functions $A$, $G$, and $P_g$ are known and appropriate numerical fluxes have been chosen.
%\remarkA{we shouldn't mention the method yet and therefore not the fluxes}
%App;
%pressure block coefficients, Ass; non-pressure block coefficients
%(saturation block(2) or saturation and concentration
%block(3)) and Aps; Asp; the coupling coefficients:
%
For Model A (i.e. \eqref{syst:1presswetnonwetsat}-\eqref{syst:2presswetnonwetsat}), we have for example $p=p_w,s=s_n$ and
\begin{alignat*}{2}
  A_{pp}(s) &= (\lambda_n(s)+\lambda_w(s))\mathbb{K}~,\qquad &
  A_{ps}(s) &= \lambda_n(s) p'_c(s) \mathbb{K}, \\
  A_{sp}(s) &= \lambda_n(s)\mathbb{K}~,\qquad &
  A_{ss}(s) &= \lambda_n(s) p'_c(s) \mathbb{K}, \\
     G_s(s)    &= 0,  & G_p(s)    &= - (\rho_w\lambda_w(s) + \rho_n \lambda_n(s)) \mathbb{K} \mathbf{g}, \\
  P_g &= \rho_n \mathbf{g},  & \\
    q_p &= q_w+q_n,  & q_s &= q_n.
\end{alignat*}

%%%%%%%%%%%%%%%%%%%%%%%%%%%%%%%%%%%%%

%\section{Discretization}
%
%\subsection{Discontinuous Galerkin Method}
%
%% explain how to generate a generic DG scheme based on the PDE form given
%% in previous section. At the moment a nonsymmetric DG version is
%% implemented and I'm not quite sure how to make that symmetric based on
%% the diffusiveFLux/source functions? Would be nice to add that generically
%% if possible otherwise I would keep it in the version it is now and ignore
%% the symmetrization terms

\section{Discretization}
\label{sec:discretization}

 In this section, we provide a discretization framework for a two-phase flow in a strongly heterogeneous and anisotropic porous medium.
\subsection{Space Discretization}
Let $\mathcal{T}_h = \{E\}$ be a family of non-degenerate, quasi-uniform, possibly non-conforming partitions of $\Omega$ consisting of $N_h$ elements (quadrilaterals or triangles in 2d, tetrahedrons or hexahedrons in 3d) of maximum diameter $h$. Let $\Gamma^{h}$ be the union of the open sets that coincide with internal interfaces of elements of $\mathcal{T}_h$. Dirichlet and Neumann boundary interfaces are collected in the set $\Gamma^{h}_{D}$ and $\Gamma^{h}_{N}$. Let $e$ denote an interface in $\Gamma^{h}$ shared by two elements $E_{-}$ and $E_{+}$ of $\mathcal{T}_h$; we associate with $e$ a unit normal vector $\nu_e$ directed from $E_{-}$ to $E_{+}$.
We also denote by $\abs{e}$ the measure of $e$. The discontinuous finite element space is $\mathcal{D}_r(\mathcal{T}_h)= \{ v \in \mathbb{L}^2(\Omega) : v_{\mid E} \in \mathcal{P}_{r_E}(E) \hspace{0.2cm} \forall E \in \mathcal{T}_{h} \}$, with  $r=(r_E)_{E \in \mathcal{T}_h}$, $\mathcal{P}_{r_E}(E)$ denotes  $\mathbb{Q}_{r_E}$ (resp. $\mathbb{P}_{r_E}$) the space of polynomial functions of degree at most $ r_E \ge 1$ on $E$ (resp. the space of polynomial functions of total  degree $r_E \ge 1$ on $E$). We approximate the pressure and the saturation by discontinuous polynomials of total degrees $r_p=(r_{p,E})_{E\in \mathcal{T}_h} $ and $r_s=(r_{s,E})_{E\in \mathcal{T}_h}$ respectively.\\
For any function $q\in \mathcal{D}_r(\mathcal{T}_h) $, we define the jump operator $ \llbracket \cdot \rrbracket $ and the average operator $\{ \cdot \}$ over the interface $e$:

$\forall e \in \Gamma^{h}$, \hspace{5pt}  $\llbracket q \rrbracket :=  q_{E_{-}} \nu_{e}  -  q_{E_{+}} \nu_{e}$,\hspace{5pt}    $\{ q \} :=  \frac{1}{2} q_{E_{-}}  +  \frac{1}{2} q_{E_{+}}$,

$\forall e \in \partial \Omega$,  \hspace{5pt}  $\llbracket q \rrbracket :=  q_{E_{-}}\nu_{}$,  \hspace{5pt} $\{ q \} :=  q_{E_{-}} $.\\
 In order to treat the strong heterogeneity of the permeability tensor, we follow~\cite{ern2010discontinuous} and introduce a weighted average operator $ \{ \cdot \}_\omega$:

 $\forall e \in \Gamma^{h},  \hspace{5pt} \{ q\}_\omega= \omega_{E_{-}} q_{E_{-}} +\omega_{E_{+}}  q_{E_{+}}$,
 
 $\forall e \in \partial \Omega,  \hspace{5pt} \{ q\}_\omega= q_{E_{-}} $.\\
 The weights are  $\omega_{E_{-}}= \frac{k^{+}}{k^{+}+k^{-}}$, \hspace{5pt} $  \omega_{E_{+}}= \frac{k^{-}}{k^{+} +k^{-}}$
with $k^{-}= \nu_e^T K_{E_{-}} \nu_e$ and $k^{+}= \nu_e^T K_{E_{+}} \nu_e$. Here, $K_{E_{-}}$ and $K_{E_{+}}$ are the permeability tensors for the elements $E_{-}$ and $E_{+}$.

% explain how to generate a generic DG scheme based on the PDE form given
% in previous section. At the moment a nonsymmetric DG version is
% implemented and I'm not quite sure how to make that symmetric based on
% the diffusiveFLux/source functions? Would be nice to add that generically
% if possible otherwise I would keep it in the version it is now and ignore
% the symmetrization terms
 The derivation of the semi-discrete DG formulation is standard (see \cite{bastian2014fully}, \cite{ern2010discontinuous}, \cite{klieber2006adaptive}). 
First, we multiply each equation of \eqref{eq:1generalFramework}-\eqref{eq:2generalFramework} by a test function and integrate over each element, then we apply Green formula to obtain the semi-discrete weak DG formulation. 
%
%Hence, the semi-discrete formulation consists in finding  the continuous in time approximations $p_{w,h}(\cdot,t) \in \mathcal{D}_{r_{p}}(\mathcal{T}_h)$, $s_{n,h}(\cdot,t) \in  \mathcal{D}_{r_{s}}(\mathcal{T}_h)$.
%
%We use a DG method to discretize this equation in space.
%
 The bulk integrals are thus given by:
\begin{align*}
  B_p( (p,s),\varphi; \bars) &=
           \sum_{E \in \mathcal{T}_h}\int_E \big( A_{pp}(\bars)\nabla p + A_{ps}(\bars)\nabla s
                                   \big) \cdot\nabla\varphi
                    +  \sum_{E \in \mathcal{T}_h} \int_E G_p(\bar s) \cdot\nabla\varphi 
                    -  \sum_{E \in \mathcal{T}_h} \int_E q_p  \varphi \\
  B_s( (p,s),\varphi; \bars) &=
            \sum_{E \in \mathcal{T}_h} \int_E \big( A_{sp}(\bars)(\nabla p-P_q) + A_{ss}(\bars)\nabla s
                                   \big) \cdot\nabla\varphi
 					  +  \sum_{E \in \mathcal{T}_h} \int_E G_s(\bar s) \cdot\nabla\varphi 
                    -  \sum_{E \in \mathcal{T}_h} \int_E q_s  \varphi \\\end{align*}
The consistency terms on the skeleton are
\begin{align*}
  C_p( (p,s),\varphi ;\bars) &= \sum_{e \in \Gamma^{h}_{} \cup \Gamma^{h}_{D} \cup \Gamma^{h}_{N}} \int_e \wavg{ A_{pp}(\bars)\nabla p + A_{ps}(\bars)\nabla s + G_p(\bar s) }
                                   \cdot \jump{\varphi}, \\
  C_s( (p,s),\varphi; \bars) &= \sum_{e \in \Gamma^{h}_{} \cup \Gamma^{h}_{D} \cup \Gamma^{h}_{N}} \int_e \wavg{ A_{sp}(\bars)(\nabla p-P_q) + A_{ss}(\bars)\nabla s + G_s(\bar s) }
                                   \cdot \jump{\varphi}.
\end{align*}
%where the weighted averages are defined by
%\begin{eqnarray*}
%  \wavg{u} := \omega^- u^+ + \omega^+ u^-
%\end{eqnarray*}
%with weights $\omega^+ = \nu^+\cdot \mathbb{K}^+\nu^+$ and $\omega^+ = \nu^-\cdot \mathbb{K}^-\nu^-$.
%
To stabilize the scheme we define interior penalty terms on the skeleton
with $\sigma>0$ a given constant:
\begin{align*}
  S_p( p,\varphi) &=  \sigma \sum_{e \in \Gamma^{h}_{} \cup \Gamma^{h}_{D} } \int_e \gamma_e^p 
                                   \jump{p} \cdot \jump{\varphi}, \\
  S_s( s,\varphi) &= \sigma \sum_{e \in \Gamma^{h}_{} \cup \Gamma^{h}_{D}} \int_e \gamma_e^s
                                   \jump{s} \cdot \jump{\varphi}.
\end{align*}
We follow the suggestions from \cite{ainsworth:09} and choose $\sigma = r (r + 1 )$ where $r$ is the highest polynomial degree of the discrete spaces.
The penalty terms $\delta_p$ and $\delta_s$ depend on the largest eigenvalues of $A_{pp}(0.5)$ and $A_{ss}(0.5)$, respectively. For Model A they are given by 
\begin{alignat*}{2}
  \gamma_e^p &=   \max(\delta_p^+,\delta_p^-) \frac{2 k^{+} k^{-}}{k^{+} +k^{-}} \times \frac{|e|}{\min(|E_+|,|E_-|)},\\
  \gamma_e^s &=   \max(\delta_s^+,\delta_s^-) \frac{2 k^{+} k^{-}}{k^{+} +k^{-}} \times \frac{|e|}{\min(|E_+|,|E_-|)},
\end{alignat*}
where
%the largest Eigenvalues of
%$A_{pp}(0.5)$ and $A_{ss}(0.5)$, respectively. For example in 2d with
%symmetric $\mathbb{K}$ satisfying $k_{11}=k_{22}$
%the penalty terms for Model A are
\begin{alignat*}{2}
  \delta_p &=  ( l_n(0.5)+l_w(0.5) ) \quad \mbox{and} \quad &
  \delta_s &=   l_n(0.5)p'_c(0.5).
\end{alignat*}
%\begin{alignat*}{2}
%  \delta_p &= (k_{11}+|k_{12}|) ( l_n(0.5)+l_w(0.5) ),\qquad &
%  \delta_s &= (k_{11}+|k_{12}|) l_n(0.5)p'_c(0.5)
%\end{alignat*}
%\remarkA{At the moment we are using the Ern formulation based on
%$ k=n\cdot Kn$}
The two bilinear forms thus are
\begin{align*}
  F_p( (p,s),\varphi ; \bars ) &= B_p( (p,s),\varphi ; \bars ) -
                                  C_p( (p,s),\varphi ; \bars ) + S_p( p,\varphi ), \\
  F_s( (p,s),\varphi ; \bars ) &= B_s( (p,s),\varphi ; \bars ) -
                                  C_s( (p,s),\varphi ; \bars ) + S_s( p,\varphi ).
\end{align*}

%\subsection{Time stepping}

% start with theta method and then explain the different iteration
% approaches to get from u^n -> u^{n+1}, i.e., impes, implicit...
\subsection{Time stepping}
\label{sec:timestepping}

% start with theta method and then explain the different iteration
% approaches to get from u^n -> u^{n+1}, i.e., impes, implicit...

%\subsection{Temporal Discretization}

Denoting with $(p^i,s^i)$ the approximation to the solution in the discrete
function space at some point in
time $t^i$ we use a simple one step scheme to
advance the solution $(p^i,s^i)$ at time $t^i$ to $(p^{i+1},s^{i+1})$ at
the next point in time $t^{i+1}=t^i+\tau$ based on
\begin{align}
  F_p( (p^{i+1},s^{i+1}),\varphi ; \bars ) &= 0, \\
  \int (s^{i+1}-s^i)\varphi + \tau F^\alpha_s( (p^{i+1},s^{i+1}),\varphi ; \bars ) &= 0,
\end{align}
defining for a given constant $\alpha \in [0,1]$ the bilinear form
\begin{align}
  F_s^\alpha( (p,s),\varphi ; \bars ) &=
       (1-\alpha) F_s( (p^i,s^i),\varphi ; s^i ) +
           \alpha F_s( (p^{i+1},s^{i+1}),\varphi ; \bars ).
\end{align}
The starting point of the iteration $(p^0,s^0)$ are taken as an $L^2$
projection of the functions given by the initial conditions into the
discrete space.
In our tests we have always used $\alpha=1$ since we were more interested
in investigating the influence of $\bars$.
We also used a fixed time step
$\tau$ throughout the whole course of the simulation although varying time steps
can be easily used as well.

Different choices for $\bars$ lead to different approaches for handling the
nonlinearities in the pressure. We tested five different approaches
described in the following:

\paragraph{Linear}
    For this approach we simply take $\bars=s^i$ leading to a
    forward Euler time stepping scheme.
\paragraph{Implicit}
    Taking $\bars=s^{i+1}$ leads to a backward Euler scheme.
    The resulting fully coupled system is solved iteratively using a
    Newton method.
\paragraph{Iterative}
    This is similar to the previous approach, replacing the Newton method
    by an outer fixed point iteration to solve the system:
    we define $\bars^k = s^{i+1,k}$ with $\bars^0=s^i$
    and in each step of the iteration we therefore solve for $k\geq 0$:
    \begin{align}
      F_p( (p^{i+1,k+1},s^{i+1,k+1}),\varphi ; s^{i+1,k} ) &= 0, \\
      \int (s^{i+1,k+1}-s^i)\varphi +
         \tau F^\alpha_s( (p^{i+1,k+1},s^{i+1,k+1}),\varphi ; s^{i+1,k} ) &= 0.
    \end{align}
\paragraph{IMPES-iterative}
    This results in an iterative scheme using an IMPES approach.
    This is similar to the previous approach except that in each step
    of the iteration we solve
    \begin{align}
      F_p( (p^{i+1,k+1},s^{i+1,k}),\varphi ; s^{i+1,k} ) &= 0, \\
      \int (s^{i+1,k+1}-s^i)\varphi +
         \tau F^\alpha_s( (p^{i+1,k+1},s^{i+1,k+1}),\varphi ; s^{i+1,k} ) &= 0.
    \end{align}
\paragraph{IMPES}
    Finally we use a classical IMPES approach, which is similar to the previous without carrying out the iteration: the saturation in the pressure equation is taken explicitly and the new pressure is used in the saturation equation (in contrast to the first approach where the old pressure is used). 
       \begin{align}
         F_p( (p^{i+1},s^{i}),\varphi ; s^{i} ) &= 0, \\
         \int (s^{i+1}-s^i)\varphi +
            \tau F^\alpha_s( (p^{i+1},s^{i+1}),\varphi ; s^{i} ) &= 0.
       \end{align}

With the exception of the first and the last approach, all
methods use an iteration to obtain a fixed point to the fully implicit
equation
\begin{align}
  F_p( (p^{i+1},s^{i+1}),\varphi ; s^{i+1} ) &= 0, \\
  \int (s^{i+1}-s^i)\varphi + \tau F^\alpha_s( (p^{i+1},s^{i+1}),\varphi ; s^{i+1} ) &= 0.
\end{align}
In the third and the fourth method this is achieved using an outer iteration
(based on the first or the last method, respectively) while the second
method uses a Newton method. To make the approaches easier to compare, we use the same stopping criteria for the iteration in all three cases.
We take $(p^{i+1},s^{i+1})=(p^{i+1,l},s^{i+1,l})$ with $l$ such that
\begin{equation} \|s^{i+1,l}-s^{i+1,l-1}\|_{L^2(\Omega)} <
      {\rm tol}_{\rm iter} \|s^{l-1}\|_{L^2(\Omega)}~.
\end{equation}
We use a value of ${\rm tol}_{\rm iter}=3\cdot 10^{-2}$ to stop the iteration when the relative change between two steps is less then three percent.

%\subsection{Adaptivity}

% provide the residual estimate and the algorithm for doing hp adaptivity
\subsection{Adaptivity}
\label{sec:adaptivity}
Different adaptive strategies are possible depending on how elements are refined/coarsened; 
whether the elements should be p-refined or h-refined; 
when should the refinement process be stopped (e.g. maximum level of refinement, stopping criterion). 
Keeping this in focus, we provide in this section a brief 
introduction to different adaptive strategies implemented 
and tested in this work. In all that follows, 
the parameters $maxpoldeg$ and $maxlevel$ refer respectively 
to the maximum polynomial degree and the maximum level of refinement allowed.

\subsubsection{Error indicator}
In the sequel, we implement an explicit estimator originally designed for non-steady 
convection-diffusion problems. A thorough analysis is available in \cite{sun20052}.

 Applying the estimator to the phase conservation equation \eqref{eq:2generalFramework} yields:
% \todo[inline]{In the python code, we computed all the results  with r=1 (in the estimator formula)}
%
\begin{align}
  \label{eq:estimator}
\eta_{E}^{2}=& h^{2}_{E}\norm{R_{vol}}^{2}_{L^{2}(E)}+\frac{1}{2}\sum\limits_{e \in \Gamma^{h}}\left( \ h_e \norm{R_{e_2}}^{2}_{L^{2}(e)}+\frac{1}{h_e}\norm{R_{e_1}}^{2}_{L^{2}(e)} \ \right) \nonumber \\
&+ \sum\limits_{e \in \partial E \cap \partial \Omega}\left( \ h_e \norm{R_{e_2}}^{2}_{L^{2}(e)}+\frac{1}{h_e} \norm{R_{e_1}}^{2}_{L^{2}(e)} \ \right).
\end{align}
Here $R_{vol}$ is the interior residual indicating  how accurate the discretized solution satisfies the original PDE at every interior point of the domain,
\begin{align*}
R_{vol} =  q_{s} - \phi \frac{ \partial s_{} }{\partial t}+ \nabla \cdot \biggl( A_{sp}(s)(\nabla p-P_g) + A_{ss}(s)\nabla s + G_s(s) \biggr).
\end{align*} 
\\
 The term $R_{e_{1}}$ is the numerical zero order inter-element (resp. Dirichlet boundary condition) residual depending on the jump of the discrete solution at the elements boundaries (resp. at the Dirichlet boundary), hence reflecting the regularity of the DG approximation (resp. the accuracy of the approximation on the Dirichlet boundary),

\begin{align*}
  R_{e_{1}} =\left\{
  \begin{array}{@{}ll@{}}
   \sigma \gamma_e^s  \llbracket s\rrbracket  & \text{if}\ e\in \Gamma^{h}, \\
     \sigma \gamma_e^s ( s_D - s)  & \text{if}\ e\in \Gamma^{D}.
  \end{array}\right.
\end{align*}
\\
 The term $  R_{e_{2}} $ is the first order numerical inter-element residual (resp. Neumann boundary condition residual) depending on the jump of numerical approximation of the normal flux at the elements boundaries (resp. at the Neumann boundary). It also allows to assess the regularity of the DG approximation (resp. the accuracy of the approximation on the Neumann boundary),
\begin{align*}
  R_{e_{2}} =\left\{
  \begin{array}{@{}ll@{}}
    \llbracket  A_{sp}(s)(\nabla p-P_g) + A_{ss}(s)\nabla s + G_s(s) \rrbracket         \cdot \nu_e & \text{if}\ e\in \Gamma^{h}, \\
     J_n + \left(  A_{sp}(s)(\nabla p-P_g) + A_{ss}(s)\nabla s + G_s(s) \right) 			   \cdot \nu_e & \text{if}\ e\in \Gamma^N.
  \end{array}\right.
\end{align*} 

\subsubsection{Adaptive strategies}
The indicator presented above will be used to drive adaptive algorithms. The h-adaptive algorithm is depicted in Algorithm \ref{algo:hadaptive1}.
Given the error indicator $\eta_E^{r,n}$ defined in equation \eqref{eq:estimator}
for a polynomial degree $r$ in time step $n$ for each element $E$, 
we refine each element whose error indicator is greater than a refinement threshold value 
$hTol^n_E$ and we coarsen elements where the indicator is smaller than the coarsening threshold $0.01\times hTol^n_E$.

In order to automatically compute the tolerance for refinement $hTol^n_E$ used in each timestep 
during the simulation we choose the following approach. We pre-describe a tolerance 
for the initial adaptation such that the resulting refined grid looks
satisfactory. Then we applied an equi-distribution strategy which aims to
equally distribute the error contribution over all time steps and grid elements. 
As a result we compute $hTol^n_E$ based on the initially computed error
indicator, 
\begin{equation}
  hTol^n_E := tTol\, \frac{\tau^n}{|\mathcal{T}^n_h|} \qquad \mbox{with} \qquad   
  tTol     := \frac{1}{T}\sum_{E \in \mathcal{T}_h} \eta^{r,0}_E. 
\end{equation}
 %
% The use of heuristic error indicators requires a maximum level of allowed h-refinement $maxlevel$ to be specified to avoid overly aggressive refinement.
% 
In the following we use $\eta_E^r = \eta_E^{r,n}$ as abbreviation for ease of
reading.
The choice between increasing or decreasing the local polynomial order depends heavily on the value of an indicator $\varsigma_E (\eta_E^{r}, \eta_E^{r-1})$ where $\eta^r_E$, $E \in\mathcal{T}_h$ is a given error indicator and $\eta_E^{r-1}$ is the same indicator evaluated for the $\mathbb{L}^{2}$ projection of the solution into a lower order polynomial space. The derivation of this $\mathbb{L}^2$ projection is quite straightforward due to the hierarchical aspect of the modal DG bases implemented. 
We considered a marking strategy based on the difference of $\varsigma_E
=|\eta_E^{r}- \eta_E^{r-1}|$. When this difference on a given element is non zero we expect the higher order to contribute to the accuracy of the scheme and keep or increase the given polynomial on that element otherwise the polynomial order is decreased.
\begin{tabular}{lcr}
\begin{minipage}[t]{0.5\textwidth}
\begin{algorithm}[H]
 \caption{h-adapt}
\label{alg:h-adapt}
  \begin{algorithmic}[1]
   \State {Let $\eta_E^{r,n}$ be given}
      \ForAll{$E \in \mathcal{T}_h$} 
        \State { $h_E=diam(E)$ }
        \If {$\eta_E^{r,n}\textgreater hTol^n_E$  AND $maxlevel\textgreater level(E)$}
   			\State $h_E^{new} :=\frac{ h_E} {2}$
         \ElsIf {$\eta_E^{r,n}\textless 0.01\times hTol^n_E$ AND $level(E) > 0$}
            \State $h_E^{new} :=2 h_E$
         \Else
            \State $h_E^{new} := h_E$
       \EndIf

      \EndFor
    \end{algorithmic}\label{algo:hadaptive1}
\end{algorithm}
\end{minipage} &&
\begin{minipage}[t]{0.4\textwidth}
\begin{algorithm}[H]
 \caption{p-adapt: markpDiff}
  \label{alg:p-adapt}
    \begin{algorithmic}[1]
                  \State {Let $\varsigma_E$ be given}
      \ForAll{$E \in \mathcal{T}_h$} 
        \State { $r_E:=poldeg(E)$ }
           \If {$\varsigma_E \textless  ptol$ }
                 \If {$r_E \textgreater 1$ }
                	 		\State {$r_E^{new}:=r_E -1 $}
                 \Else	
                	 		\State {$r_E^{new}:=r_E $}
                	\EndIf
            \ElsIf {$\varsigma_E \textgreater 100 \times ptol$ }
               	 \If {$r_E \textless maxpoldeg$ }
							\State{$r_E^{new}:=r_E +1$}      
                	 \Else	
                	 		\State {$r_E^{new}:=r_E $}
                	  \EndIf
            \Else 
                  	  \State {$r_E^{new}:=r_E $}  
          \EndIf

      \EndFor
    \end{algorithmic}\label{algo:markpDiff}
 \end{algorithm}
\end{minipage} 
\end{tabular}

\subsection{Stabilization}
\label{sec:stabilization}

Although due to the presence of the capillary pressure terms strong shocks do not occur 
in the numerical experiments carried out in this paper, the DG schemes needs
stabilization to avoid unphysical values, such as negative
saturation which would lead to an undefined state 
in equation \eqref{BCcappress}.

% \subsubsection{Positivity preserving limiter}

We follow the approach from \cite{cheng:13} which has been initially proposed
by Zhang and Shu in \cite{zhang:10}. The general idea is to scale each
polynomial on each element such that a constraint on
minimum and maximum values of the saturation is respected.
We define the following projection operator $\Pi_{s}: \mathcal{D}_r(\mathcal{T}_h) \longrightarrow \mathcal{D}_r(\mathcal{T}_h)$ with 

\newcommand{\ics}{\bold{x}}
\newcommand{\stabfactor}{\chi}

\begin{equation}
  \int_{\Omega} \Pi_s[s] \cdot \varphi  =  \int_{\Omega} \tilde{s} \cdot \varphi
  \quad \forall \varphi \in \mathcal{D}_r 
\end{equation}
where on each element $E$ of the grid we define a scaled saturation 
$\tilde{s}(\ics) := \stabfactor_E\big( s(\ics) - \bar{s}\big) + \bar{s}$ with 
$\bar{s}$ being the mean value of $s$ on element $E$.
The scaling factor is  
\begin{equation}
  \stabfactor_E := \min_{\ics \in \Lambda_E}\{ 1, |(\bar{s} - s_{min})/(\bar{s} - s(\ics))|, |(s_{max} - \bar{s})/(\bar{s} - s(\ics))| \}
\end{equation}
for the combined set of all quadrature points $\Lambda_E$ used for evaluation of the bilinear
forms defined earlier, i.e. interior and surface integrals.
  
The scaling limiter is applied after each Newton iteration for the implicit
scheme and after each iteration of the iterative schemes.

\section{Numerical Experiments}
\label{sec:results}
This section provides different numerical experiments aiming to demonstrate
the effectiveness and robustness of the DG discretization of porous media flow
models. All test are implemented with the hp-adaptive DG method described
in the previous section using the different approaches for the time step
computation. The maximal grid level was fixed to three and the
maximal polynomial was also three. The main components of the code areddescribed
in some detail in \ref{sec:code} and provide as part of a docker container
as explained in \ref{sec:docker}.
We also show results based on some alternative
approaches for example for the underlying model or for the adaptive
strategy. These modifications to the python code are also provided in
Appendix~\ref{sec:code}. They demonstrate the flexibility of the Python
code.

\subsection{Problem setting}\label{sec:problemsetting}
%A vertical DNAPL infiltration Flow over a low permeability lens
A container is filled with two kinds of sand and saturated by water with density $\rho_w=1000 \ Kg/m^{3}$  and viscosity $\mu_w=1\times10^{-3} \ Kg/m \ s$. The dense non-aqueous phase liquid (DNAPL) considered in the experiment is Tetrachloroethylene with density $\rho_n=1460 \ Kg/m^{3}$ and viscosity $\mu_n=9\times10^{-4} \ Kg/m \ s$. 
%
% We set $r_p=r_s$ for all test cases. The maximal polynomial orders employed for the 2d problem are $r_p=r_s=3$.
% Although it is possible to use higher polynomial orders, the schemes become computationally expensive in terms of both storage and CPU time for practical use.
% The grids are locally adapted in a nonconforming fashion. 

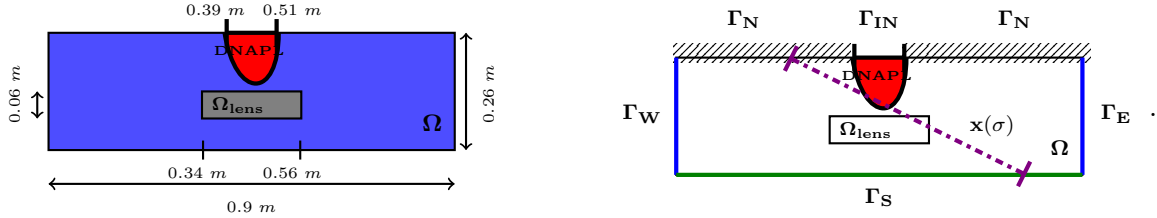
\begin{figure}[h!]
\begin{minipage}[c]{.45\linewidth}
  \begin{tikzpicture} [scale = 0.6,thick]
  \draw[fill=blue!70!white] (0,3.9) -- (9,3.9) -- (9,6.5)--(0,6.5) -- cycle;

  \draw[fill=white!50!black] (3.4,4.6) -- (5.6,4.6) -- (5.6,5.2)--(3.4,5.2) -- cycle;

  \draw[sloped,ultra thick,fill=red] (3.9,6.5) .. controls (4.14375,5.) and (5.1,5.) .. (5.1,6.5) ;
  \draw(4.44375,6.10) node[rotate=0]{{\fontsize{2}{2}\selectfont $\bf DNAPL$}} ;

  \draw[<->] (9.25,3.9) -- (9.25,6.5)  ;
  \draw(9.75,5.15) node[rotate=90]{{\fontsize{6}{6}\selectfont $0.26$ $m$}} ;

  \draw[<->] (-0.25,4.6) -- (-0.25,5.2)  ;
  \draw(-0.75,5.15) node[rotate=90]{{\fontsize{6}{6}\selectfont $0.06$ $m$}} ;

  \draw[<->] (0.00,3.15) -- (9.0,3.15)  ;
  \draw(4.5,2.65) node[rotate=0]{{\fontsize{6}{6}\selectfont $0.9$ $m$}} ;

  \draw (8.5,4.5) node {{\fontsize{9}{9}\selectfont $\bf \Omega$}};

  \draw (4.2,4.9) node {{\fontsize{7}{7}\selectfont $\bf \Omega_{ lens}$}};

  \draw[|-|, ultra thick] (3.9,6.5) -- (5.1,6.5)  ;
  \draw(3.7,7) node[rotate=0]{{\fontsize{6}{6}\selectfont $0.39$ $m$}} ;
  \draw(5.4,7) node[rotate=0] {{\fontsize{6}{6}\selectfont $0.51$ $m$}} ;

  %\draw[|-|] (0,4.6) -- (0,5.2)  ;
  %\draw(-1.5,5.2) node[rotate=0] {{\fontsize{6}{6}\selectfont $0.52$ $m$}} ;
  %\draw(-1.5,4.65) node[rotate=0] {{\fontsize{6}{6}\selectfont $0.465$ $m$}} ;

  \draw[|-|] (3.4,3.9) -- (5.6,3.9)  ;
  \draw(3.3,3.4) node[rotate=0]{{\fontsize{6}{6}\selectfont $0.34$ $m$}} ;
  \draw(5.4,3.4) node[rotate=0]{{\fontsize{6}{6}\selectfont $0.56$ $m$}} ;

  \end{tikzpicture}

 \end{minipage} 
 \hfill
  \begin{minipage}[c]{.45\linewidth}
    \begin{tikzpicture} [scale = 0.6,thick]
      \tikzstyle{ground}=[ultra thick,fill,pattern=north east lines,draw=none,minimum width=67,rotate=0,minimum height=0.25]

      \tikzstyle{ground1}=[ultra thick,fill,pattern=north east lines,draw=none,minimum width=71,rotate=0,minimum height=0.25]
      
    \draw[fill=none] (0,3.9) -- (9,3.9) -- (9,6.5)--(0,6.5) -- cycle;

    \draw[] (3.4,4.6) -- (5.6,4.6) -- (5.6,5.2)--(3.4,5.2) -- cycle;

    \node (ground) at (1.9,6.6) [ground] {};
    \node (ground1) at (7.1,6.6) [ground1] {};

    %\draw[sloped,ultra thick,fill=red] (3.9,6.5) .. controls (4.14375,5.) and (4.5625,5.) .. (4.8,6.5) ;
    %\draw(4.44375,6.10) node[rotate=0]{{\fontsize{2}{2}\selectfont $\bf DNAPL$}} ;
    \draw[sloped,ultra thick,fill=red] (3.9,6.5) .. controls (4.14375,5.) and (5.1,5.) .. (5.1,6.5) ;
    \draw(4.44375,6.10) node[rotate=0]{{\fontsize{2}{2}\selectfont $\bf DNAPL$}} ;

    \draw[ultra thick,draw=blue] (9.,3.9) -- (9.,6.5)  ;
    \draw(9.75,5.25) node[rotate=0]{{\fontsize{8}{8}\selectfont $\bf \Gamma_E$}} ;

    \draw[ultra thick,draw=blue] (0.,3.9) -- (0.,6.5)  ;
    \draw(-0.75,5.25) node[rotate=0]{{\fontsize{8}{8}\selectfont $\bf \Gamma_W$}} ;

    \draw[ultra thick,draw=green!50!black] (0.,3.9) -- (9.0,3.9)  ;
    \draw(4.5,3.4) node[rotate=0]{{\fontsize{8}{8}\selectfont $\bf \Gamma_S$}} ;

    \draw (8.5,4.5) node {{\fontsize{8}{8}\selectfont $\bf \Omega$}};

    \draw (4.2,4.9) node {{\fontsize{7}{7}\selectfont $\bf \Omega_{ lens}$}};

    \draw[|-|, ultra thick] (3.9,6.5) -- (5.1,6.5)  ;
    \draw(7.5,7.35) node[rotate=0]{{\fontsize{8}{8}\selectfont $\bf \Gamma_N$}} ;
    \draw(1.5,7.35) node[rotate=0]{{\fontsize{8}{8}\selectfont $\bf \Gamma_N$}} ;

    \draw(4.5,7.35) node[rotate=0]{{\fontsize{8}{8}\selectfont $\bf \Gamma_{IN}$}} ;

    %Draw the visualization line  x(\sigma)
    \draw[|-|, ultra thick, dash dot, draw=violet] (2.5,6.5) -- (7.75,3.9);  
    \draw(7,5.) node[rotate=0]{{\fontsize{8}{8}\selectfont $\bf x(\sigma)$}} ;
    \end{tikzpicture}
   \end{minipage}
\label{fig:geometryandbndcond}.
\caption{\small Geometry and boundary conditions for the DNAPL infiltration problem. The purple line in the right picture
         is described by $x(\sigma)$ from equation \eqref{eq:plotline}.}\label{fig:domainsetting}
\end{figure}

\begin{figure}
  \includegraphics[width=0.24\textwidth]{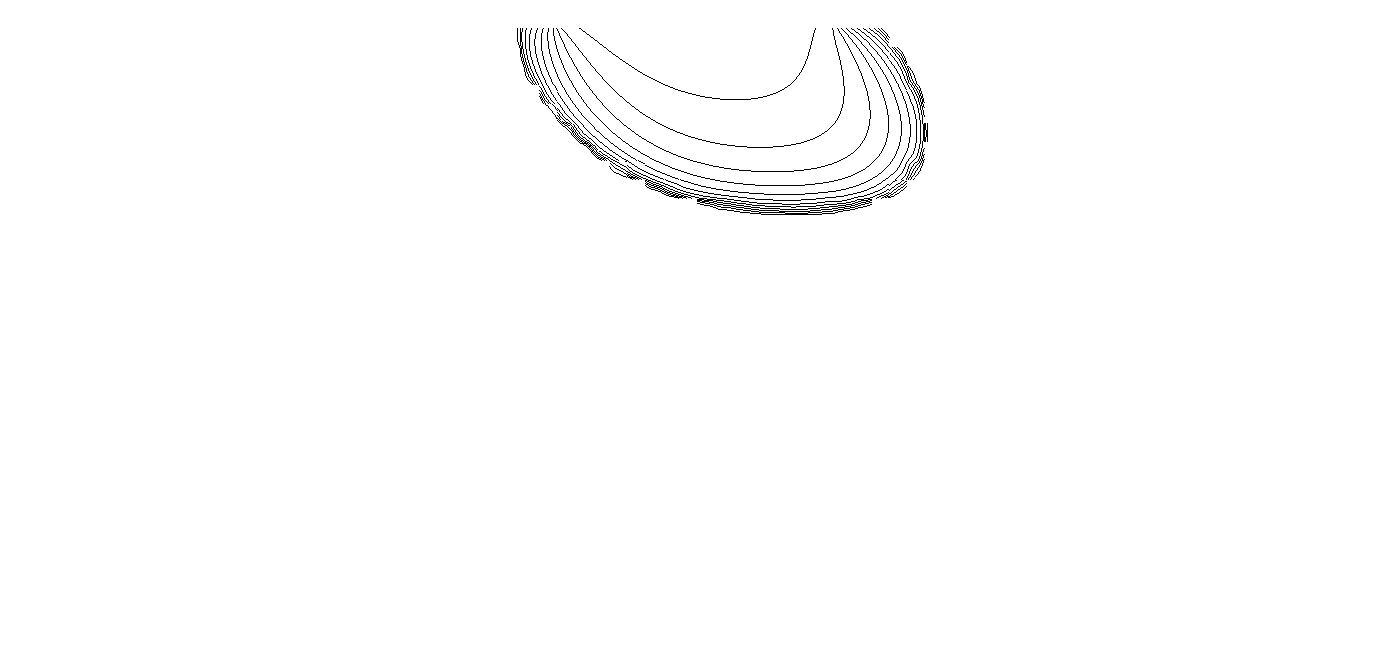}
  \includegraphics[width=0.24\textwidth]{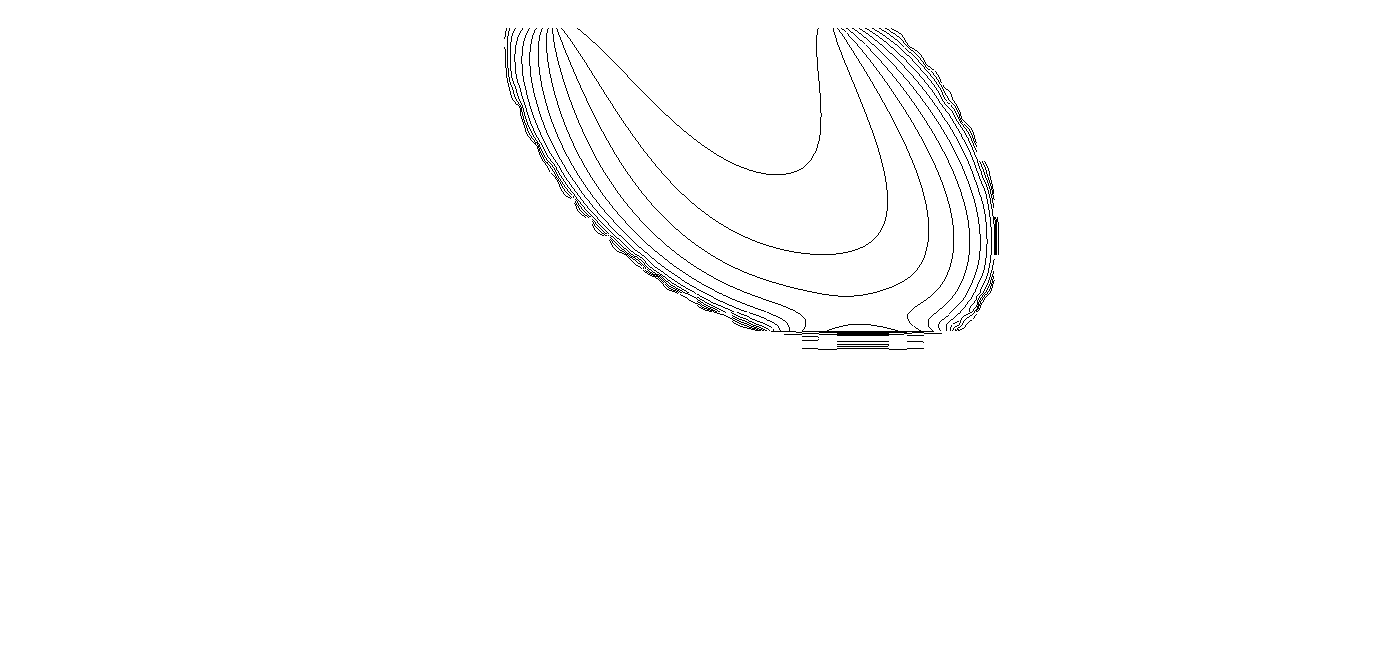}
  \includegraphics[width=0.24\textwidth]{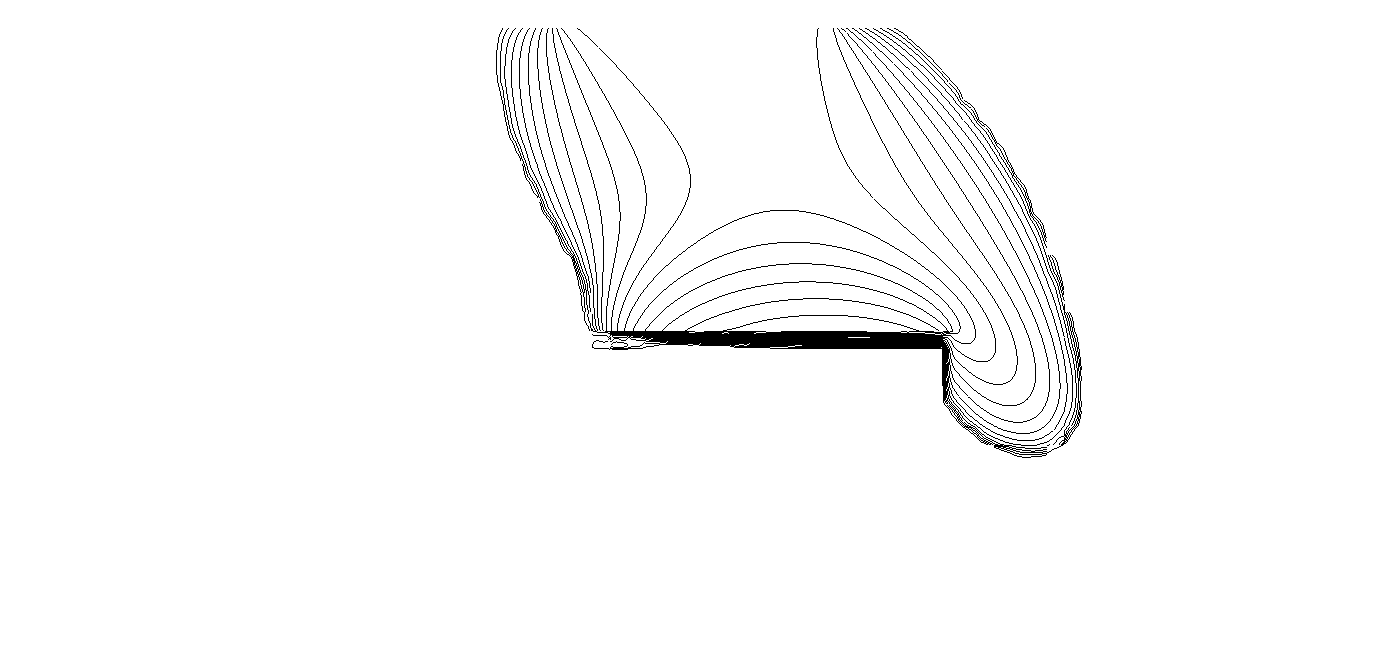}
  \includegraphics[width=0.24\textwidth]{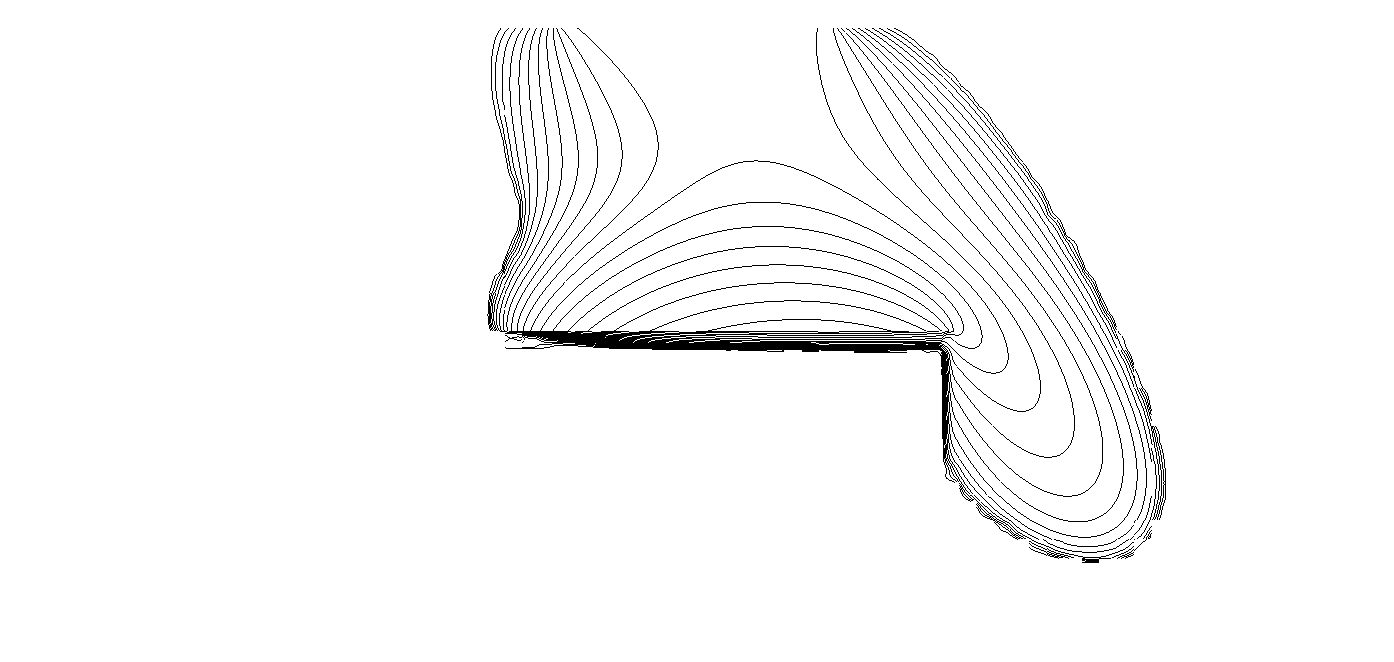}

  \includegraphics[width=0.24\textwidth]{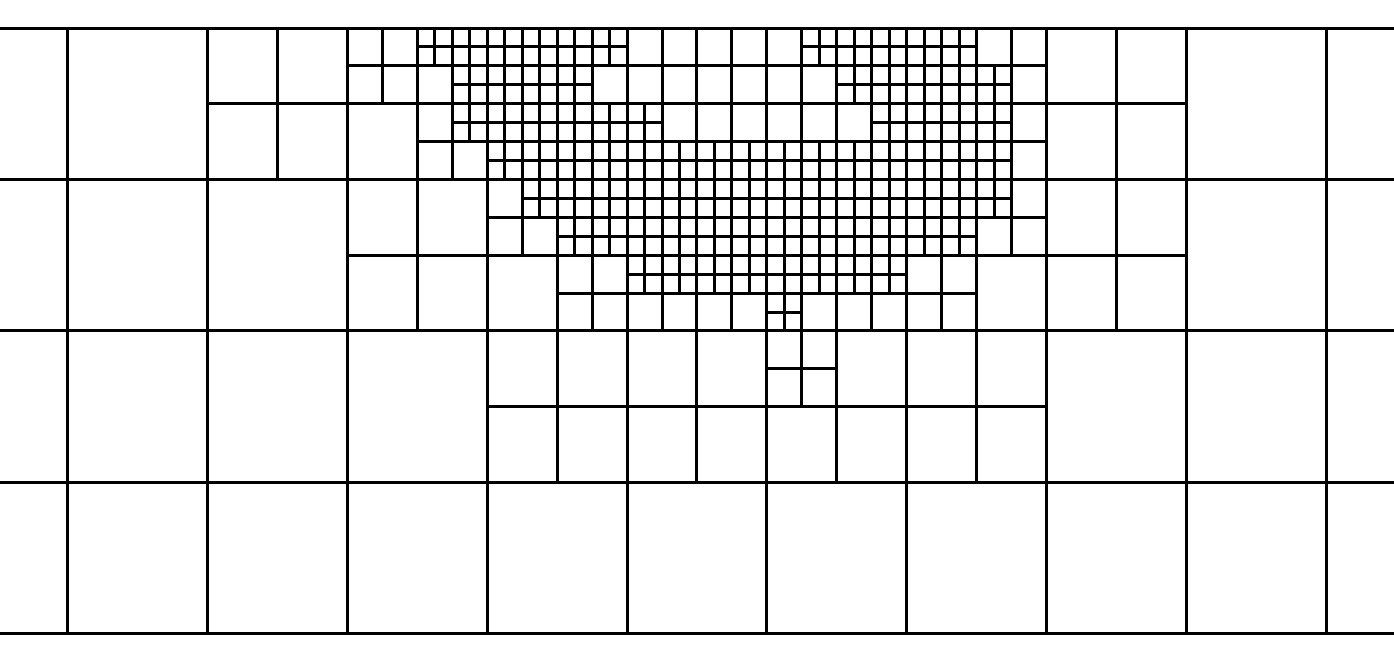}
  \includegraphics[width=0.24\textwidth]{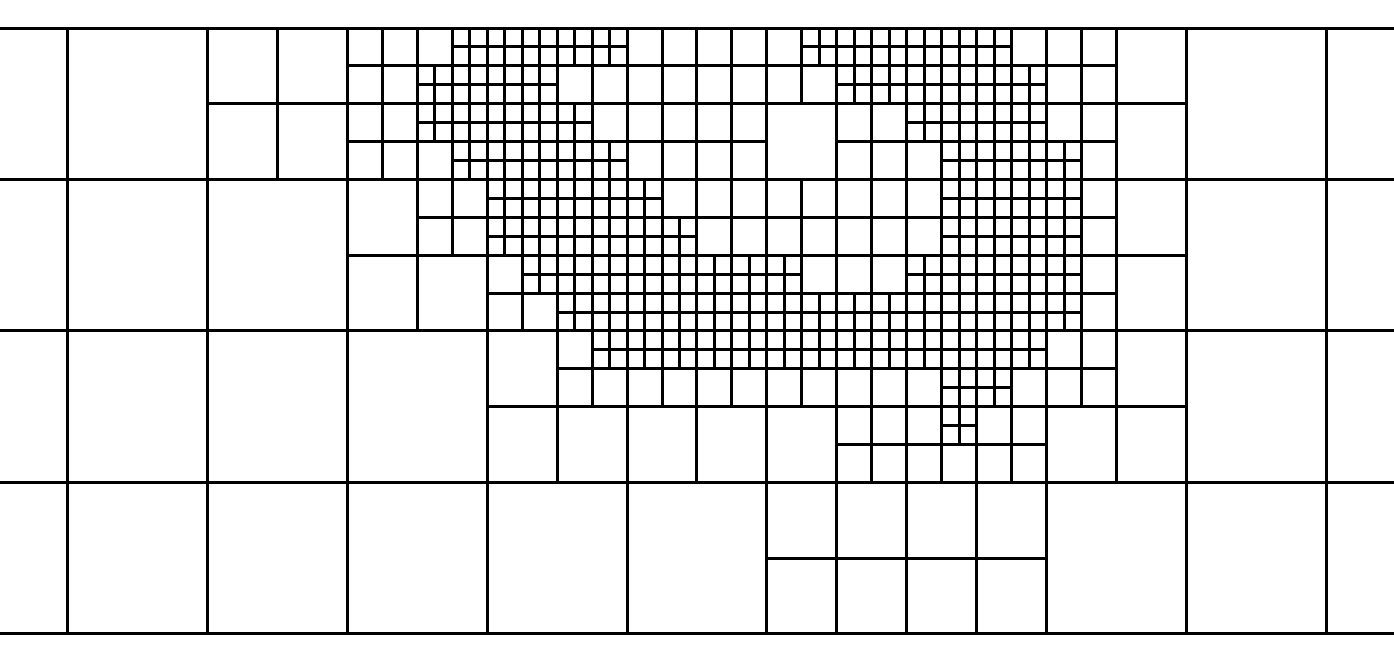}
  \includegraphics[width=0.24\textwidth]{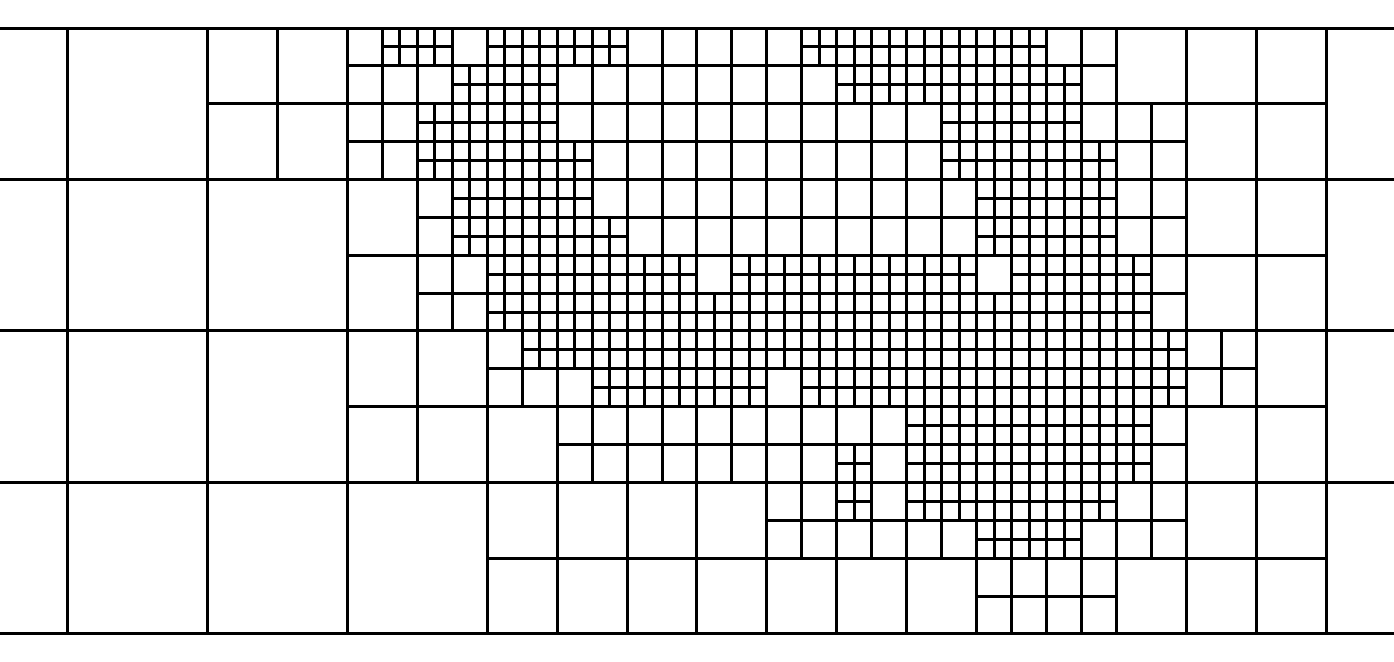}
  \includegraphics[width=0.24\textwidth]{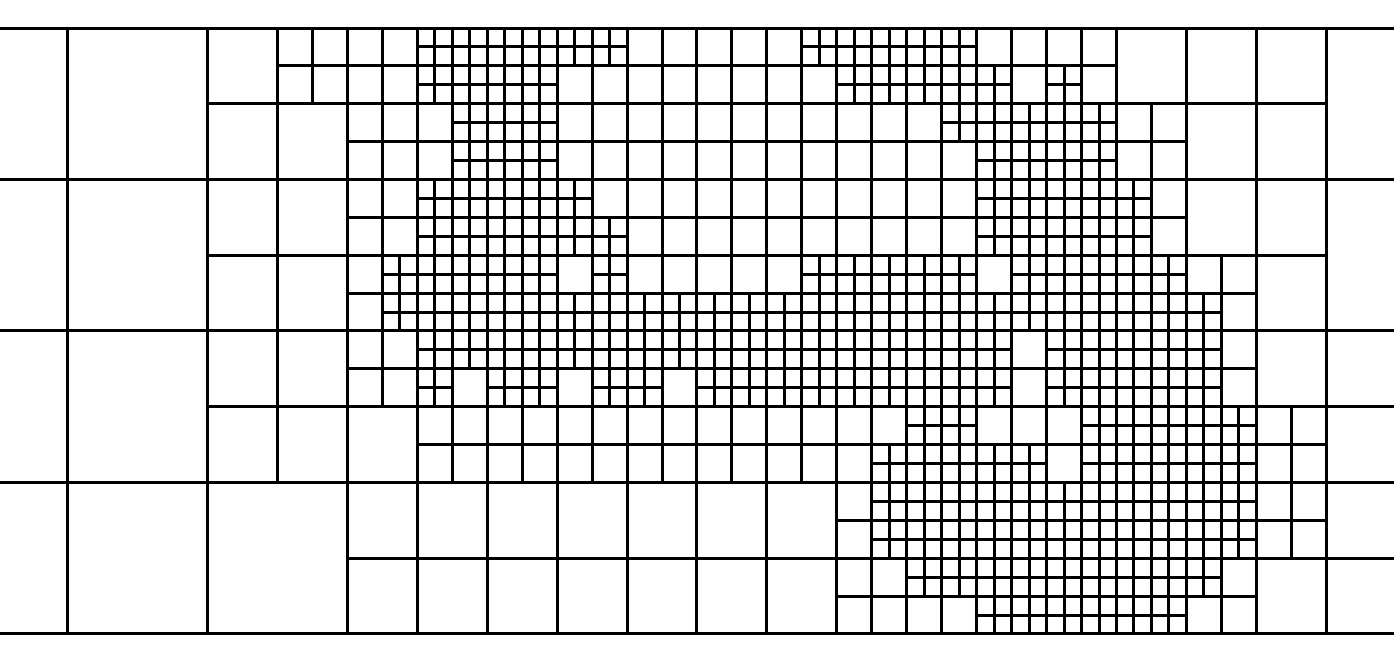}
  \caption{Evolution of the non wetting saturation $s_n$
           at times $t=200,400,600$, and $t=800$ (top) and the
           corresponding adaptive grid structure (bottom).}
  \label{fig:flowevolution}
\end{figure}

We consider a two-dimensional DNAPL infiltration problem with different sand types and anisotropic permeability tensors. The material properties are detailed in Table \ref{tab:param}. The bottom of the reservoir is impermeable for both phases. Hydrostatic conditions for the pressure $p_w$ and homogeneous Dirichlet conditions for the saturation $s_n$ are prescribed at the left and right boundaries. A flux of $J_n=-5.137\times10^{-5}  \ m\ s^{-1}$ of the DNAPL is infiltrated into the domain from the top. Detailed boundary conditions are specified in Figure \ref{fig:domainsetting} and Table \ref{tab:bndcdt}.
Initial conditions where the domain is fully saturated with water and hydrostatic pressure distribution are considered (i.e. $p_w^0=(0.65-y)\cdot9810$, $ s_n^0=0$). The permeability tensor $\mathbb{K}_{\Omega  \backslash \Omega_{lens}}$ of the domain $\Omega  \backslash \Omega_{lens}$  is
\begin{displaymath}
\mathbb{K}_{\Omega  \backslash \Omega_{lens}}=\begin{pmatrix}\label{}
10^{-10}&  -5 \times 10^{-11}\\
%& &\\
 -5 \times 10^{-11} & 10^{-10} 
\end{pmatrix}m^2
\end{displaymath}

and the permeability tensor $\mathbb{K}_{ \Omega_{lens}}$ of the lens $\Omega_{lens}$  is
\begin{displaymath}
\mathbb{K}_{ \Omega_{lens}}=\begin{pmatrix}\label{}
6 \times 10^{-14}&  0\\
%& &\\
0 & 6 \times 10^{-14} 
\end{pmatrix}m^2.
\end{displaymath}
The coarsest (macro) mesh consists of 60 quadrilateral elements globally
refined everywhere to the finest level would result in 3840 elements.
The final time is $T=800$ $ s$.
%We consider a Newton solver tolerance $newtTol=3\times10^{-7}$ and a linear solver tolerance $linabstol=2.7\times10^{-7}$. 
For visualization we later plot the solution of $s$ over the line 
\begin{align}\label{eq:plotline}
    x(\sigma) = (1 - \sigma) (0.25,0.65)^T + \sigma (0.775, 0.39)^T
\end{align}
with $\sigma \in [0,1]$.
\begin{table}[h!]
\begin{minipage}[c]{.45\linewidth}
 \scalebox {1}{ \begin{tabular}{|l|l|l|l|l|l|l|l|l|l|l|} \hline 
   $ \phantom{}$  & $\Omega_{lens}$  & $\Omega  \backslash \Omega_{lens}$  \\ \hline
  $\Phi $ [-]  & $0.39$  & $0.40$   \\ \hline
%  $k $ $[m^2]$ & $6.64\times 10^{-14}$ &   $6.64\times 10^{-11}$ \\ \hline
  $ S_{wr}$  [-]  & $0.1$  & $0.12$    \\ \hline 
$S_{nr}$ [-]  & $0.00$ & $0.00$    \\ \hline 
$\theta $ [-] & $2.0$ & $2.70$    \\ \hline 
$p_d$  [Pa]& $5000$ & $755$    \\ \hline 
  \end{tabular}}
\caption{\small 2d problem parameters.}
  \label{tab:param}
 \end{minipage} 
 \quad
  \begin{minipage}[c]{.45\linewidth}
   
    \scalebox {1}{ \begin{tabular}{|l|l|l|l|l|l|l|l|l|l|l|}\hline
  $\Gamma_{IN}$  & $J_n=-5.137\times10^{-5}$,  $J_w=0$   \\ \hline
   $\Gamma_{N}$ &$J_n=0.00$,  $J_w=0.00$ \\ \hline
   $\Gamma_{S}$  & $J_w=0$, $J_n=0.00$     \\ \hline 
  $\Gamma_{E} \cup \Gamma_{W}$  & $p_w=(0.65-y)\cdot9810$, $ s_n=0$  \\ \hline 
  \end{tabular}}
    \caption{\small 2d problem boundary conditions.}
     \label{tab:bndcdt}
   \end{minipage}
%   }
\end{table}
Snapshots of the evolution of the resulting flow and the grid structure are shown in
Figure~\ref{fig:flowevolution}.

\subsection{Time step stability}
\label{sec:timestability}

In this section we compare the various splitting and solution strategy described in 
Section \ref{sec:timestepping}. We compare three implicit and iterative
coupling schemes and two loosely coupled schemes, one of them the classical
IMPES scheme. 

\begin{figure}[!ht]
  \centering
  \includegraphics[width=0.475\textwidth]{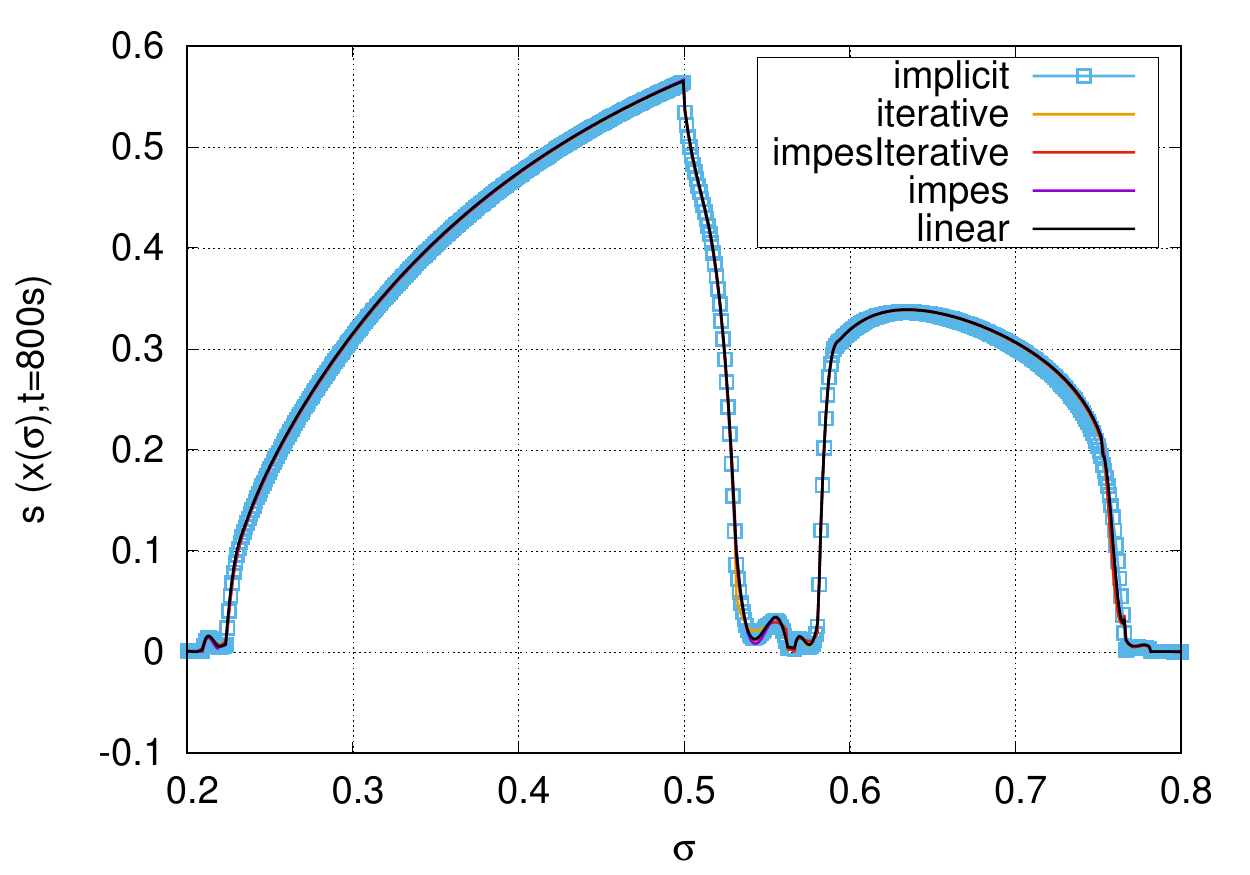}
  \includegraphics[width=0.475\textwidth]{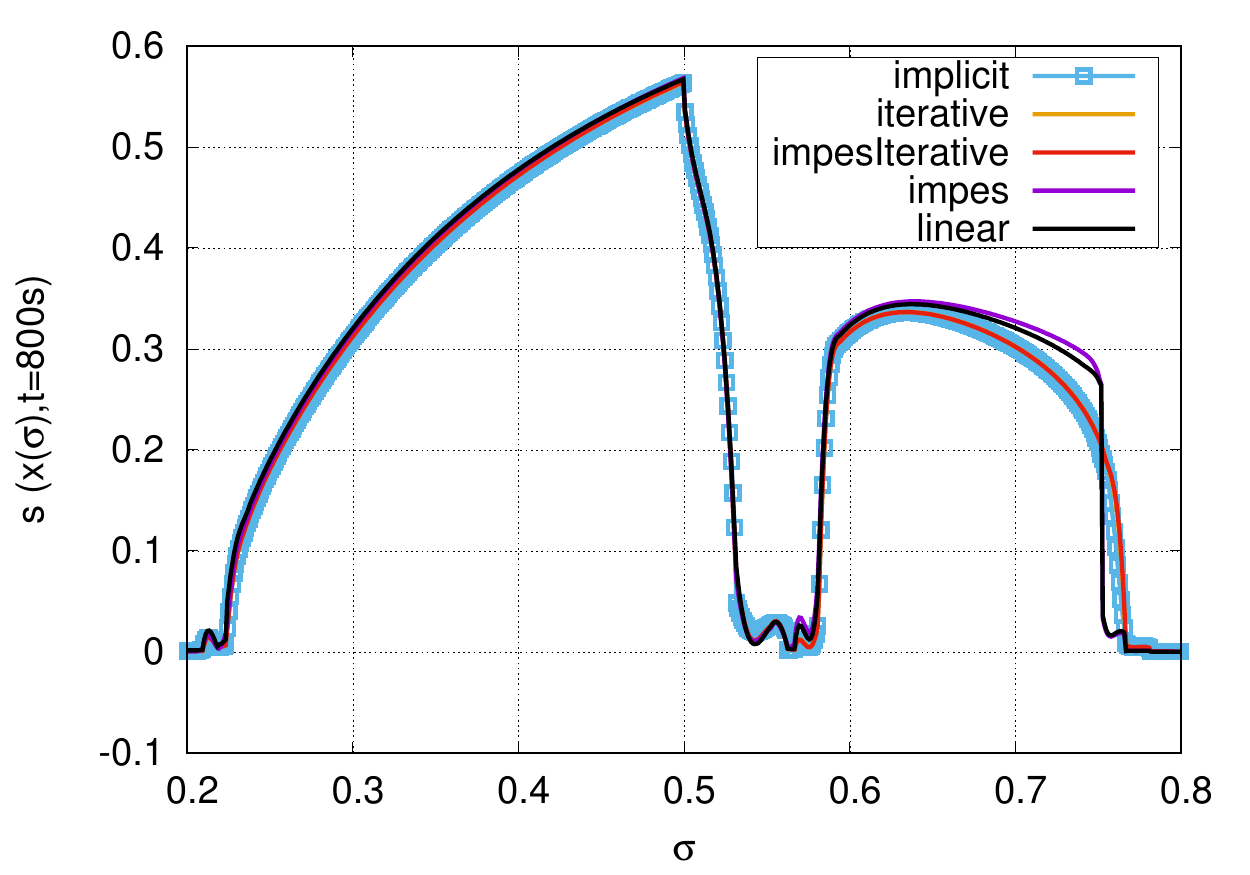}
  \caption{All schemes for $\tau=3$ (left) and $\tau = 5 $ (right). All schemes are
  able to capture the solution characteristics for times steps smaller and 
  up to $\tau=3$. For $\tau > 3$ only the fully coupled schemes 
  are able to produce reasonable solutions, while the loosely
  coupled schemes do no longer capture the front position correctly.}
  \label{fig:modelA_all}
\end{figure}

For $\tau > 3 $ only the fully coupled schemes are able to produces reasonable
solutions. This is illustrated in Figure \ref{fig:modelA_all}.

\begin{figure}[!ht]
  \centering
  \includegraphics[width=0.475\textwidth]{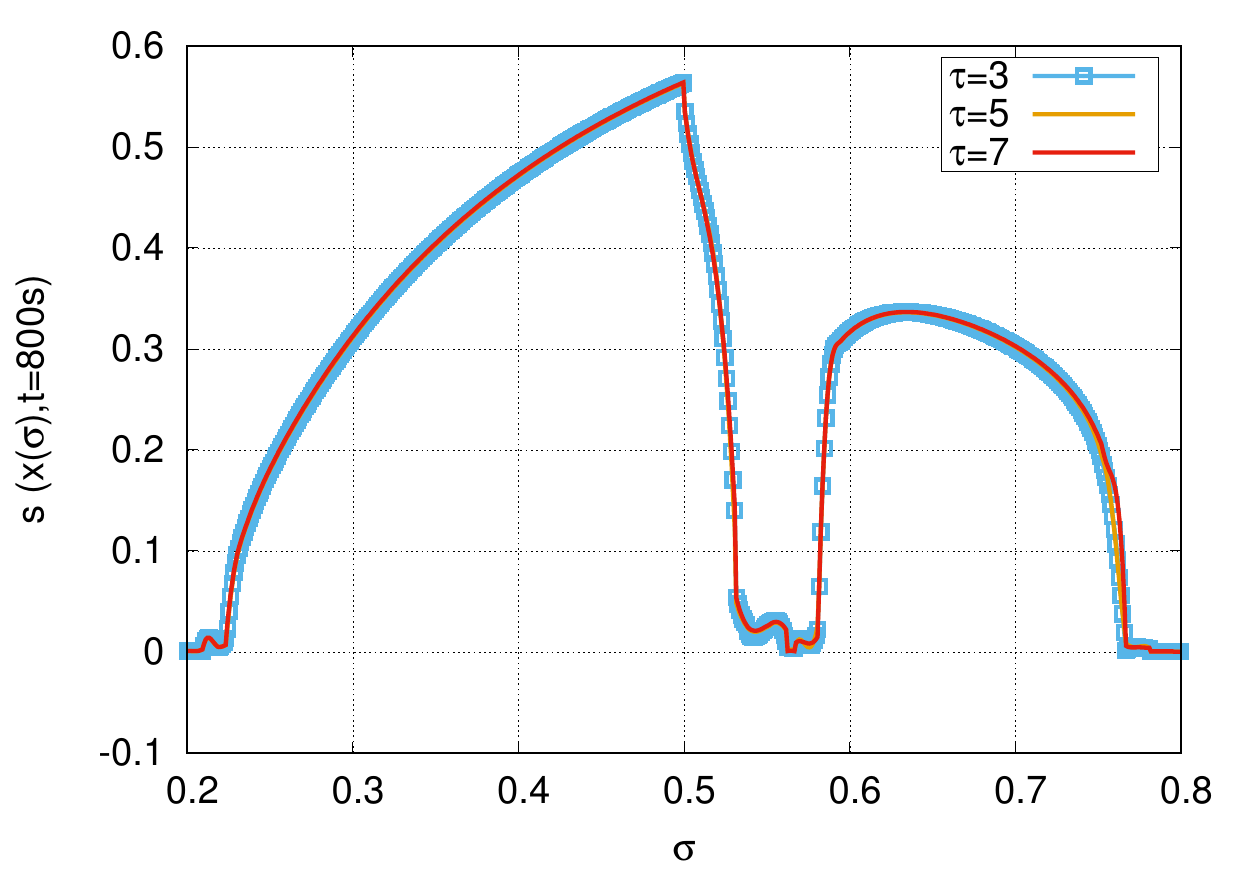}
  \includegraphics[width=0.475\textwidth]{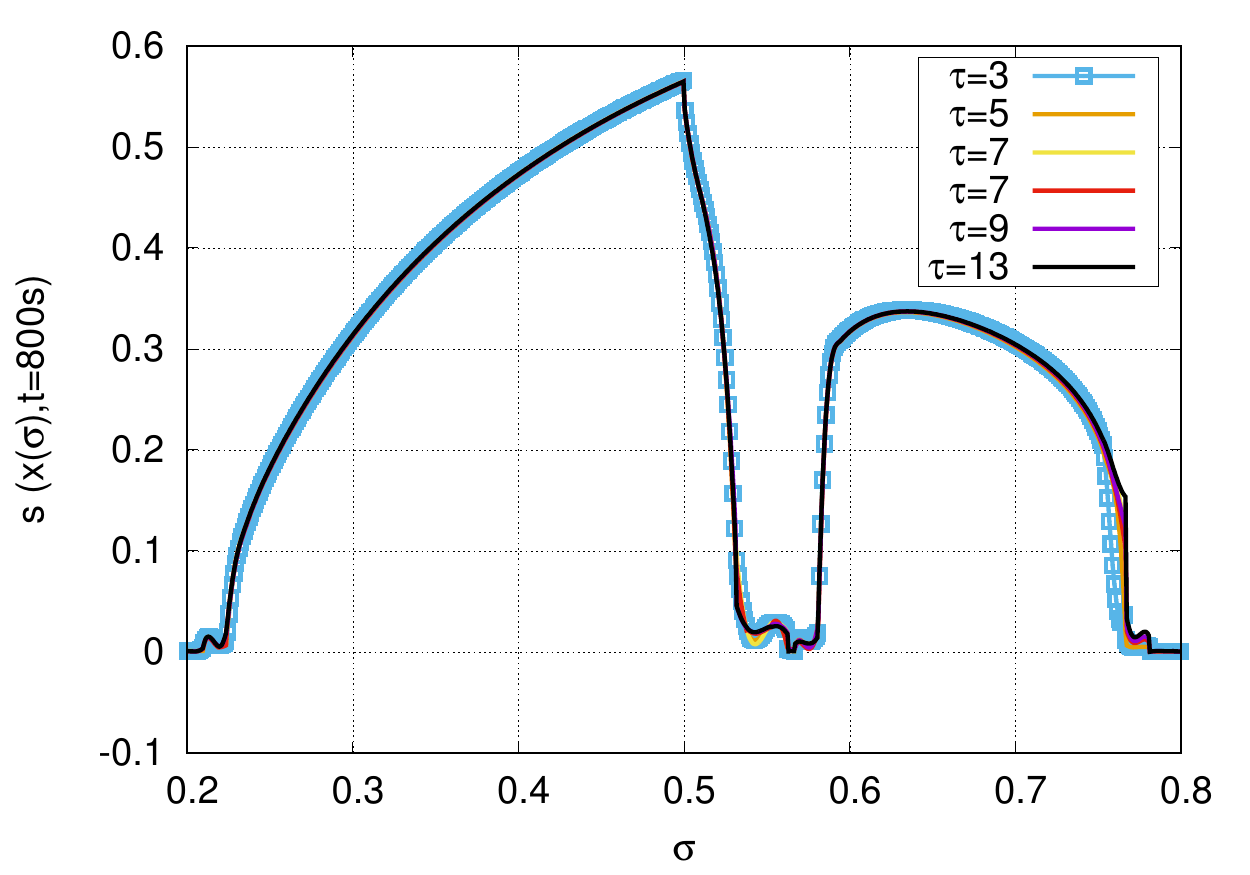}
  \caption{Left, the solution for fully coupled implicit scheme for $\tau=3,5,7$.
  Right the solution for the IMPES-iterative scheme for
  $\tau=3,5,7,9,11,13$.}
  \label{fig:modelA_dt_cmp}
\end{figure}

In Figure \ref{fig:modelA_dt_cmp} we compare the solution of the various fully coupled schemes
for different time step sizes. If the solution converges then a correct solution
profile is produced. The stability of the implicit scheme is influenced by the
fact that the stabilization operator is only applied before and after the Newton
solver. 

In principle the explicit coupling schemes work fine for small time steps and
fail to produce a valid solution for larger time steps. Here, the implicit 
schemes show their strength allowing for faster computation once the time step
is chosen sufficiently large.

\subsection{Cut off stabilization}

In this section we study a very simple stabilization approach aining at
finding a replacement for the more complicated scaling limiter described in
Section~\ref{sec:stabilization}. The idea is to simply replace values
for the saturation $s$ below a given threshold $s_{min}$ and $s_{max}$.
Values of the saturation outside of the region considered
physical will cause problems when computing the capillary pressure.
So in this approach we use a very simple
cut off for guaranteeing that no non negative values are used in the power
laws required for the capillary pressure by replacing $s_{w,e},s_{n,e}$
by $\min\{\max\{s_{w,e},\epsilon\},1-\epsilon\}$ and
$\min\{\max\{s_{w,e},\epsilon\},1-\epsilon\}$, respectively, where
$\epsilon=10^{-5}$.

This approach can be directly incorporated into the symbolic description of
the model as shown in \ref{sec:codeModifications_cufoff}.

\begin{figure}[!ht]
  \centering
  \includegraphics[width=0.32\textwidth]{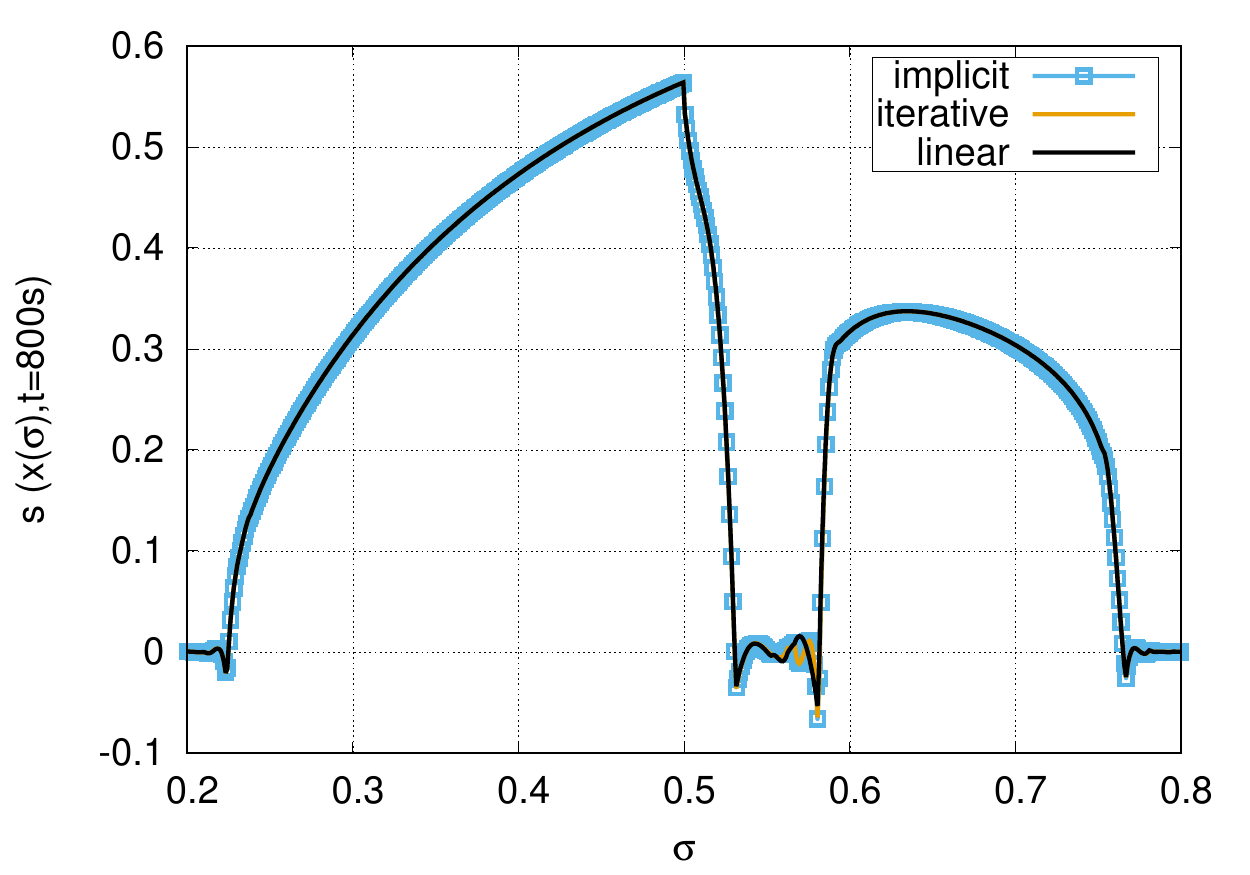}
  \includegraphics[width=0.32\textwidth]{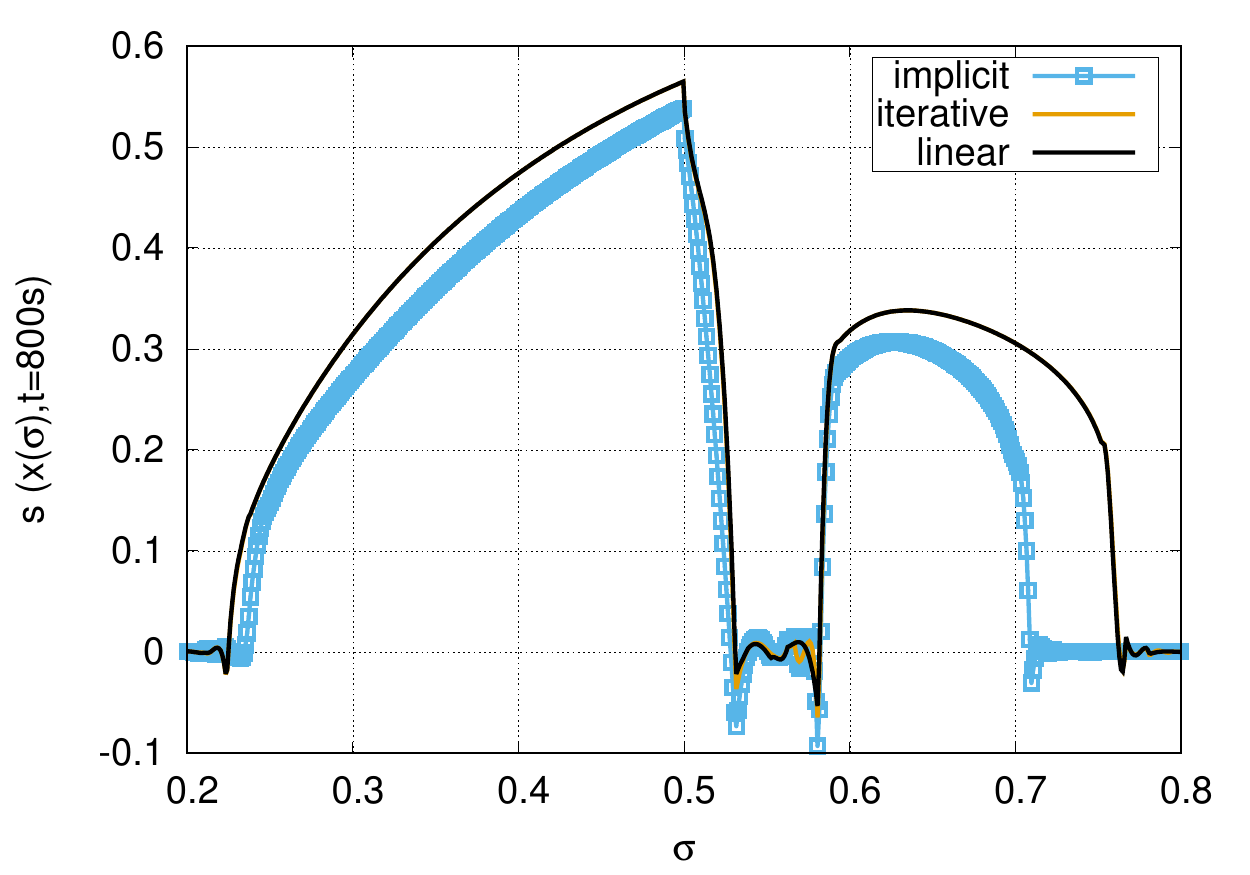}
  \includegraphics[width=0.32\textwidth]{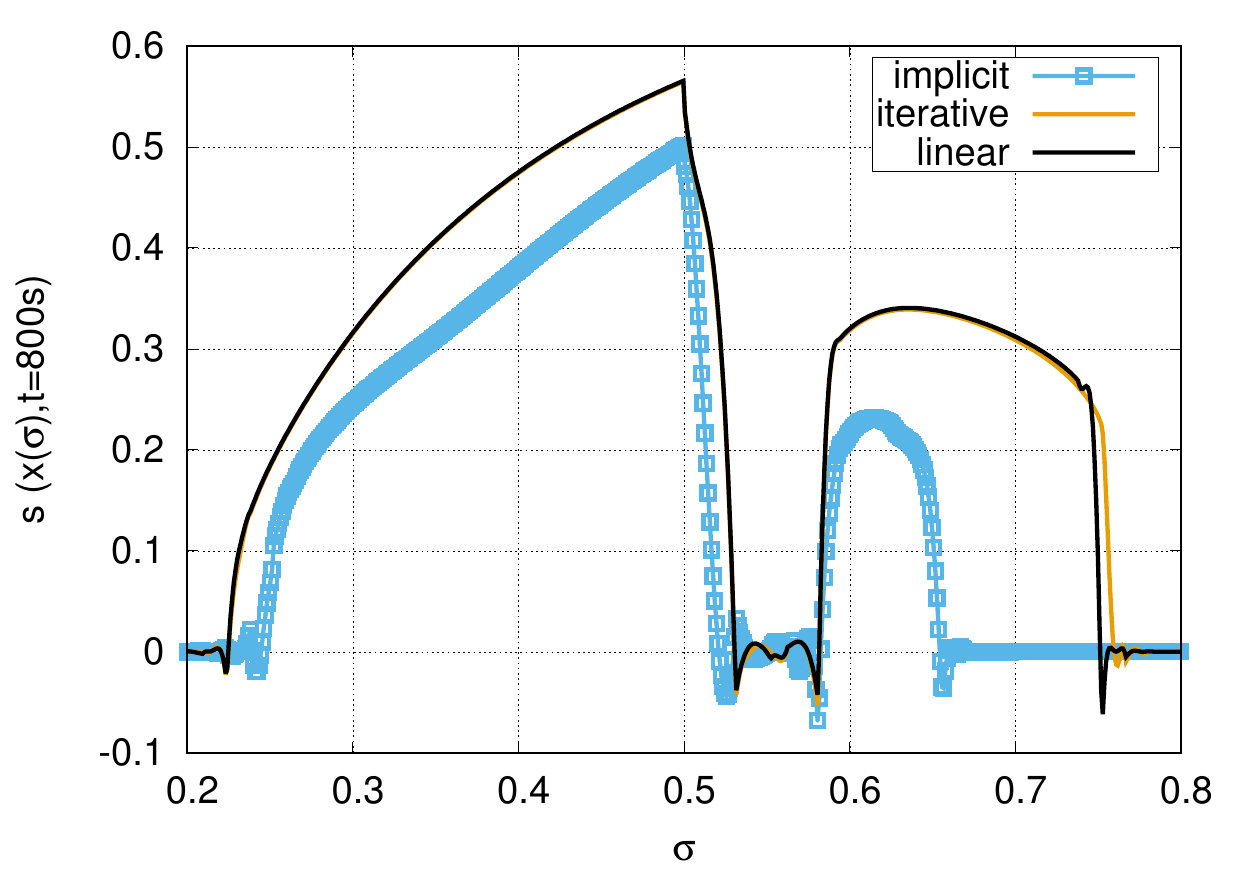}
  \label{fig:modelA_cutoff_cmp}
  \caption{Left results for $\tau = 1$, middle $\tau = 3$ and right $\tau = 5$. 
  The IMPES and impesIterative scheme fail to converge with this
  approach. The other schemes all produce oscillations around the front. 
  Interestingly, the fully coupled implicit also fails to compute the correct front position for 
  increasing times steps. }
\end{figure}

As can be clearly seen in Figure~\ref{fig:modelA_cutoff_cmp} significant
over and undershoots are produced by all methods at the fronts. Both IMPES
type splitting schemes fail to converge even for smaller time steps and the
fully coupled implicit scheme produces wrong flow speeds even for moderate
values of $\tau$. Only the iterative scheme manages to produce at least a
reasonable representation of the flow.

\subsection{Different model: Model B}

In this section we compare our original model formulation with a
description where the two-phase flow problem is modeled as a system of equations with two unknowns $\bar{p}$ and $s_n$. Here $\bar{p}=p_w+\frac{1}{2}p_c$:
\begin{align}\label{syst:1presswetnonwetsatnopc}
 -\nabla\cdot \biggl(( \lambda_w + \lambda_n ) \mathbb{K} \nabla \bar{p} + \frac{(\lambda_n- \lambda_w ) }{2}  p_c^{\prime} \mathbb{K} \nabla s_{n} - (\rho_w\lambda_w + \rho_n \lambda_n) \mathbb{K} \mathbf{g}\biggr)&= q_w + q_n \  \mbox{on} \ \Omega \times (0,T),\\
 \phi \frac{\partial  s_n}{\partial t} - \nabla \cdot \biggl( \lambda_n \mathbb{K} (\nabla \bar{p}-\rho_n \mathbf{g})\biggr) - \frac{1}{2} \nabla \cdot \biggl( \lambda_n p'_c \mathbb{K} \nabla s_{n}\biggr)  &=q_{n}\qquad \   \mbox{on} \ \Omega \times (0,T).
 \label{syst:2presswetnonwetsatnopc}
\end{align}

To complete the system, we add appropriate boundary and initial conditions. 
%\begin{align}
%s_n(\cdot,0)&= s^{0}_n(\cdot) \mbox{,} \qquad \qquad \qquad \qquad \bar{p}(\cdot,0)= p^{0}_{w}(\cdot) + \frac{1}{2} p_c(s^{0}_n(\cdot))    \qquad \quad    \mbox{in} \ \Omega, \\
% \bar{p}&=p_{w,D} + \frac{1}{2} p_c(s_D)  \mbox{,} \quad \quad   \quad \quad \quad s_n=s_{ D} \qquad  \quad  \qquad  \qquad \qquad \quad   \mbox{on} \ \Gamma^{D}_{} \times (0, T), \\
%\mathbf{v}_\alpha \cdot \nu &= J_\alpha \mbox{,} \qquad \qquad  \qquad \qquad \qquad \ J_t= \sum_{\alpha \in \{w,n\}} J_\alpha \qquad \hspace*{1.4cm} \qquad  \mbox{on}  \ \Gamma^{N}_{ } \times (0, T).
%\end{align}
\begin{alignat*}{3}
   \bar{p}(\cdot,0)&= p^{0}_{w}(\cdot) + \frac{1}{2} p_c(s^{0}_n(\cdot))  ~,\qquad &
s_n(\cdot,0)&=s^{0}_n(\cdot) ~,\qquad & \mbox{in}& \ \Omega,\\
 \bar{p}&=p_{w,D} + \frac{1}{2} p_c(s_D)   ~,\qquad &
 s_n&=s_{ D} ~,\qquad & \mbox{on}& \ \Gamma^{D}_{} \times (0, T),\\
\mathbf{v}_\alpha \cdot \nu &= J_\alpha  ~,\qquad &
 J_t&= \sum_{\alpha \in \{w,n\}} J_\alpha ~,\qquad & \mbox{on}&  \ \Gamma^{N}_{} \times (0, T).
\end{alignat*}
Here, $J_\alpha \in \mathbb{R}$, $\alpha \in \{w,n\}$ is the inflow, $s^{0}_n, \  p^{0}_w, \ s_{ D}$, and $\ p_{ w,D}$ are real numbers.

Following the general description of the problem given in Section \ref{sec:model}
we have $p=\bar p, s=s_n$ and
\begin{alignat*}{2}
  A_{pp}(s) &= (\lambda_n(s)+\lambda_w(s))\mathbb{K}~,\qquad &
  A_{ps}(s) &= \frac {\lambda_n(s) - \lambda_w(s)}{2}p'_c(s) \mathbb{K}, \\
  A_{sp}(s) &= \lambda_n(s)\mathbb{K}~,\qquad &
  A_{ss}(s) &= \frac{ \lambda_n(s)}{2} p'_c(s) \mathbb{K}, \\
    G_s(s)    &= 0,   & G_p(s)    &= - (\rho_w\lambda_w(s) + \rho_n \lambda_n(s)) \mathbb{K}\mathbf{g}, \\
  P_g &= \rho_n \mathbf{g}, & \\
      q_p &= q_w+q_n,  & q_s &= q_n. \\
\end{alignat*}
The required changes to the Python code are again minimal and described in
the \ref{sec:codeModifications_modelB}.

\begin{figure}[!ht]
  \centering
  \includegraphics[width=0.475\textwidth]{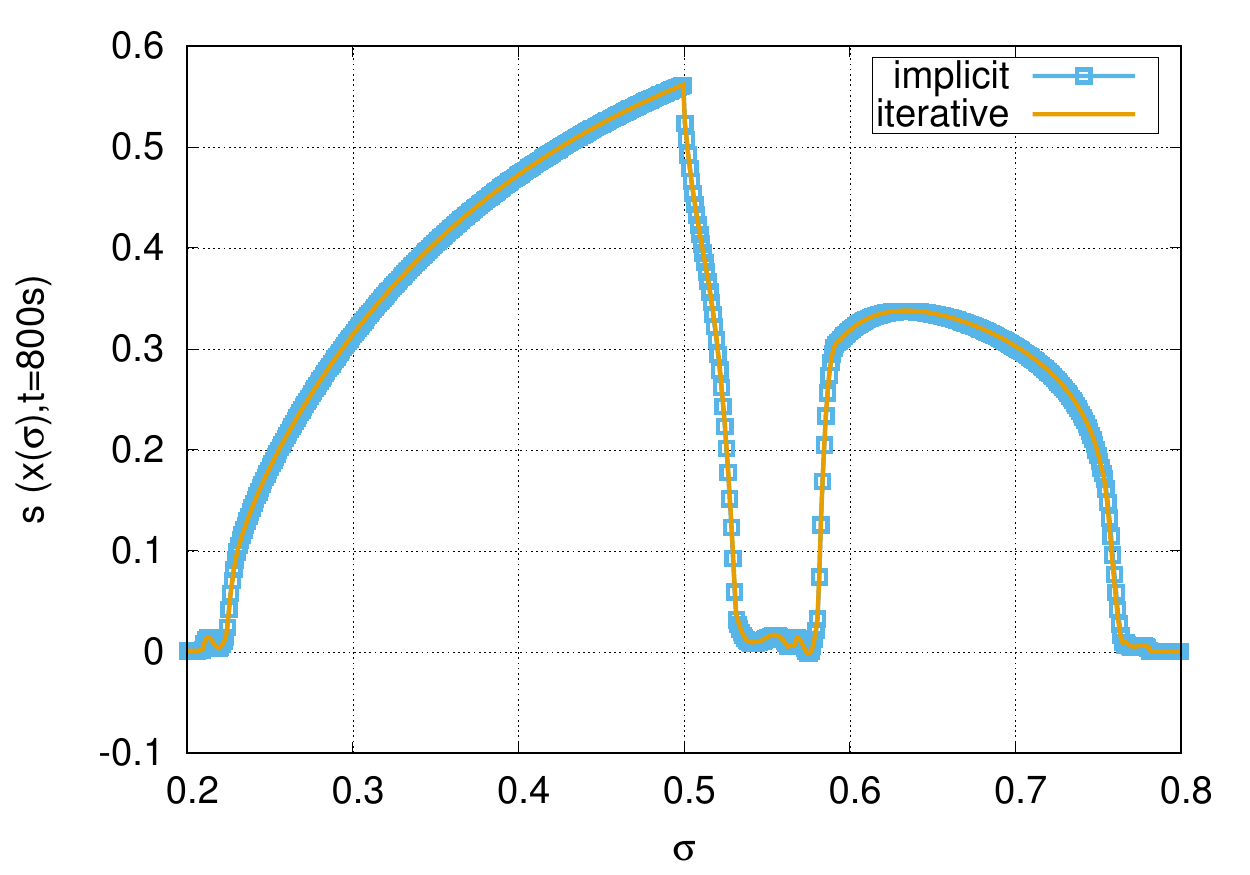}
  \includegraphics[width=0.475\textwidth]{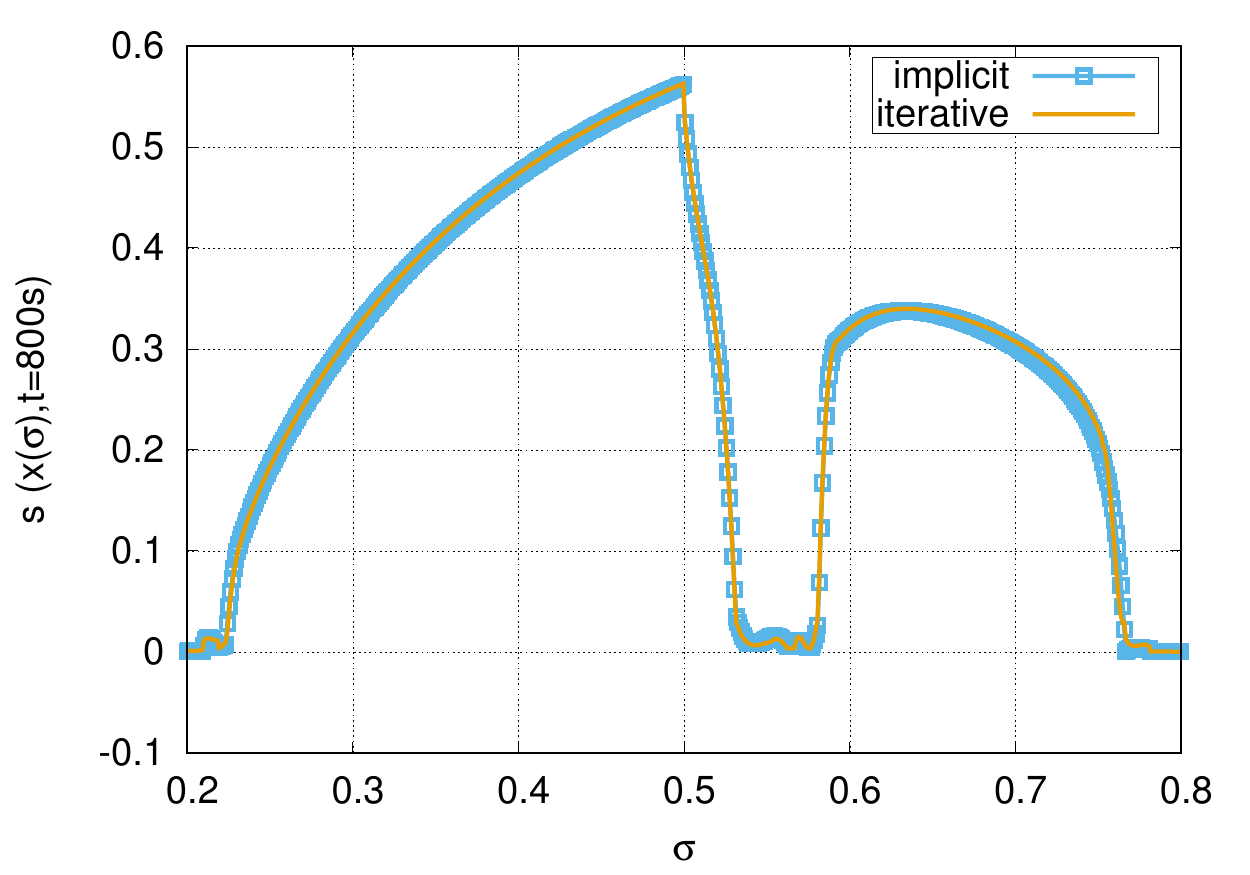}

  \centering
  \includegraphics[width=0.475\textwidth]{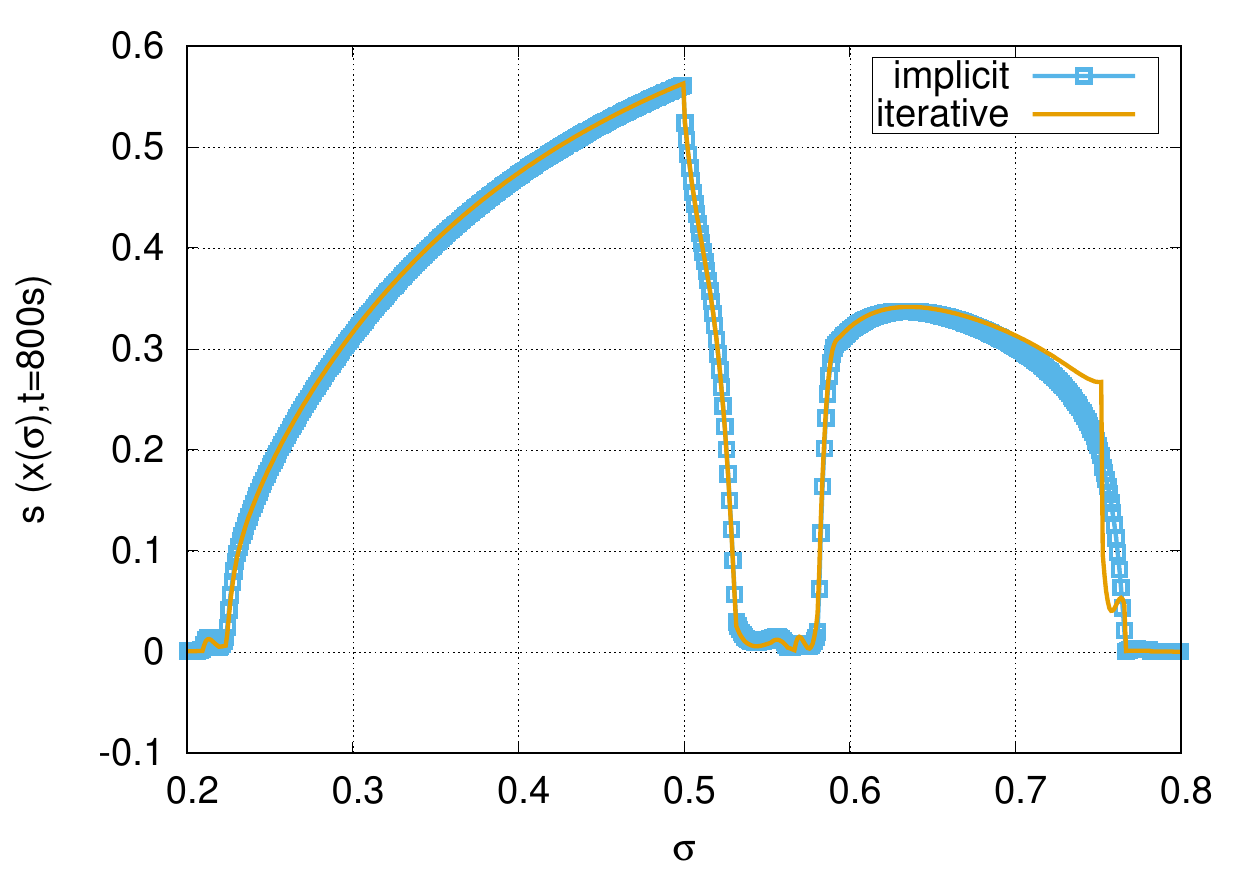}
  \includegraphics[width=0.475\textwidth]{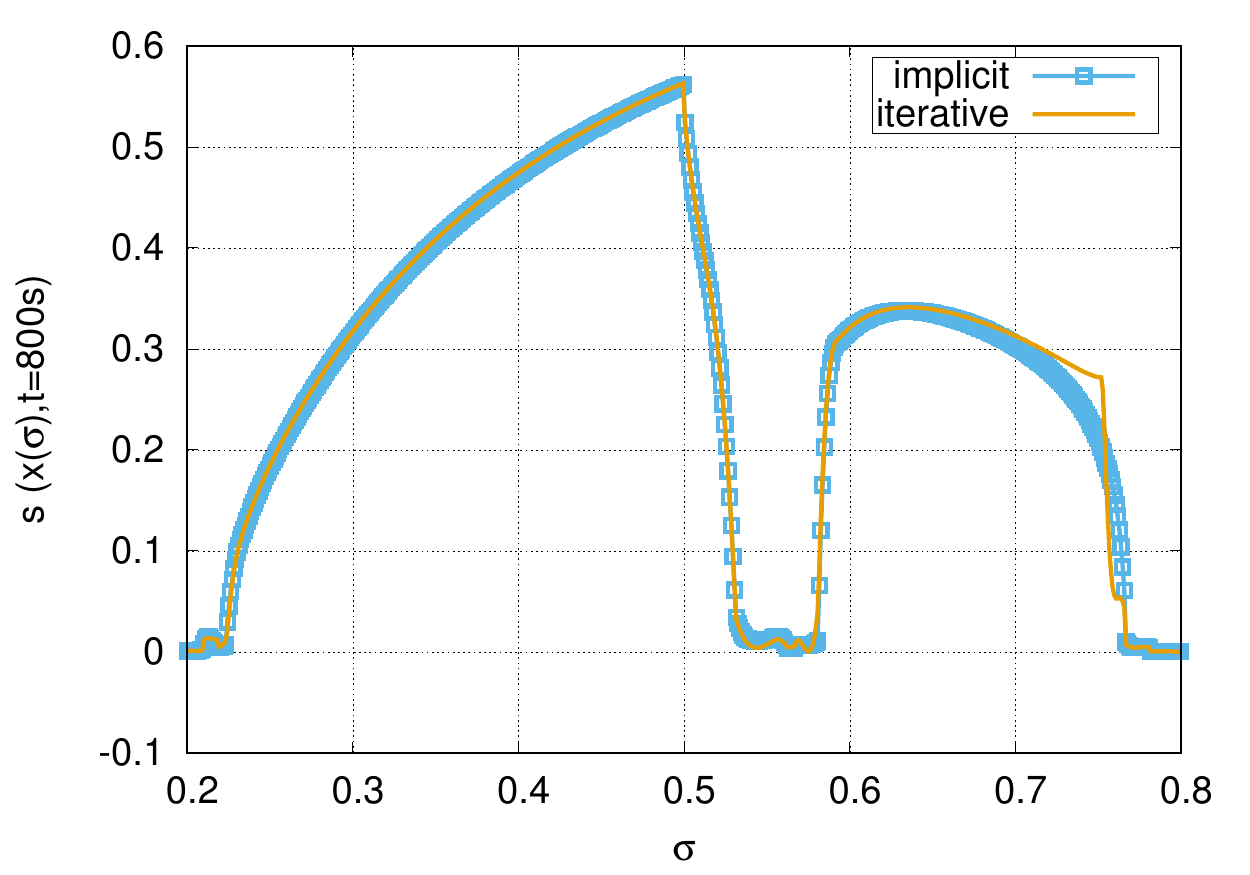}
  \caption{Results using \emph{modelB}. Top row: $\tau=1,3$, bottom row: $\tau=5,7$.}
    \label{fig:modelB_cmpdt}
\end{figure}

In the following we investigate the stability of the different methods with respect to
the time step size when applied to \emph{modelB}. We perform the same
investigation described in the previous section where we used
\emph{modelA}. The results are summarized in Figure \ref{fig:modelB_cmpdt}.
We only investigated the stability of the three methods
\emph{implicit,iterative}, and \emph{impes-iterative}. For \emph{modelB}
the splitting introduced in the \emph{impes} type approach failed even for
$\tau=1$ while the other two methods produce results in line with the
results produced with \emph{modelA} although for higher values of $\tau$
the iterative methods produce
a discontenuety at the right most front as can be seen in the plots on the
bottom row of Figure~\ref{fig:modelB_cmpdt}.  For $\tau>5$ the \emph{implicit} method
fails, making \emph{modelB} a less stable choice for this scheme. On the
other hand the \emph{iterative} approach produced results
also for larger time steps $\tau=9,11,13,15$ (not shown here)
but in each case the solution showed the same type of discontinuety.

Taking all the approaches into account, it is clear that \emph{modelA} is the more
stable representation of the problem. But our results also indicate that
the stability of the \emph{iterative} scheme does not seem to depend so
much on the choice of the model (at the least for the two versions tested) and
produces very similar results in both cases.

\subsection{P-adaptivity}

Here we compare different approaches for the indicator used to set the
local polynomial degree. In the following we always use the \emph{implicit}
method with $\tau=5$, h-adaptivity with a
maximum level of three and also a maximum level of three for the polynomial
order. In addition to the approach used previously we test a version
without p-adaptivity and an indicator based on determining the smoothness
of the solution. In regions where the indicator detects a reduction in
smoothness the polynomial order is reduced but only if the grid has been
refined to the maximum allowed level. The smoothness indicator is based on
$\varsigma_E=\frac{\eta_E^{r}}{ \eta_E^{r-1}}$ and we set $ptol=1$:
%\begin{figure}[h]
%\begin{minipage}[c]{0.45\linewidth}
         \begin{algorithm}[H]
        \caption{p-adapt: markpFrac}
\begin{algorithmic}[1]
                  \State {Let $\varsigma_E$ be given}
      \ForAll{$E \in \mathcal{T}_h$} 
        \State { $r_E:=poldeg(E)$ }
           \If {$\varsigma_E \textless  0.01 \times ptol$ }
                \If {$r_E \textless maxpoldeg$ }
              		\State {$r_E^{new}:=r_E +1 $}
                 \Else	
                	 \State {$r_E^{new}:=r_E $}
                	\EndIf
            \ElsIf {$\varsigma_E \textgreater  ptol$ }
                 \If {$r_E \textgreater 1$ }
							\State{$r_E^{new}:=r_E - 1$}      
                	 \Else	
                	 		\State {$r_E^{new}:=r_E $}
                	  \EndIf
            \Else 
                  	  \State {$r_E^{new}:=r_E $}  
          \EndIf

      \EndFor
    \end{algorithmic}\label{algo:markpFrac}
      \end{algorithm}
% \end{minipage}
%%\vspace*{-25pt}
%\end{figure}
The changes to the code are described in \ref{sec:codeModifications_padaptive}.

\begin{figure}
  \includegraphics[width=0.32\textwidth]{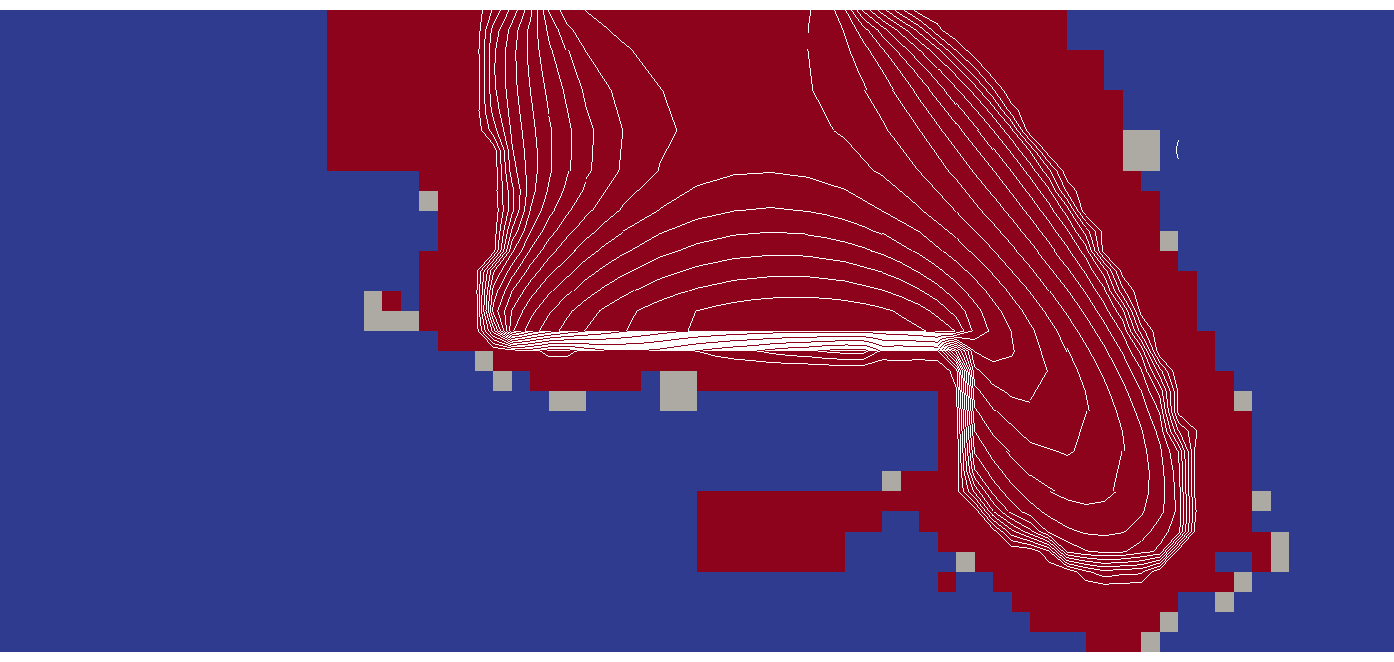}
  \includegraphics[width=0.32\textwidth]{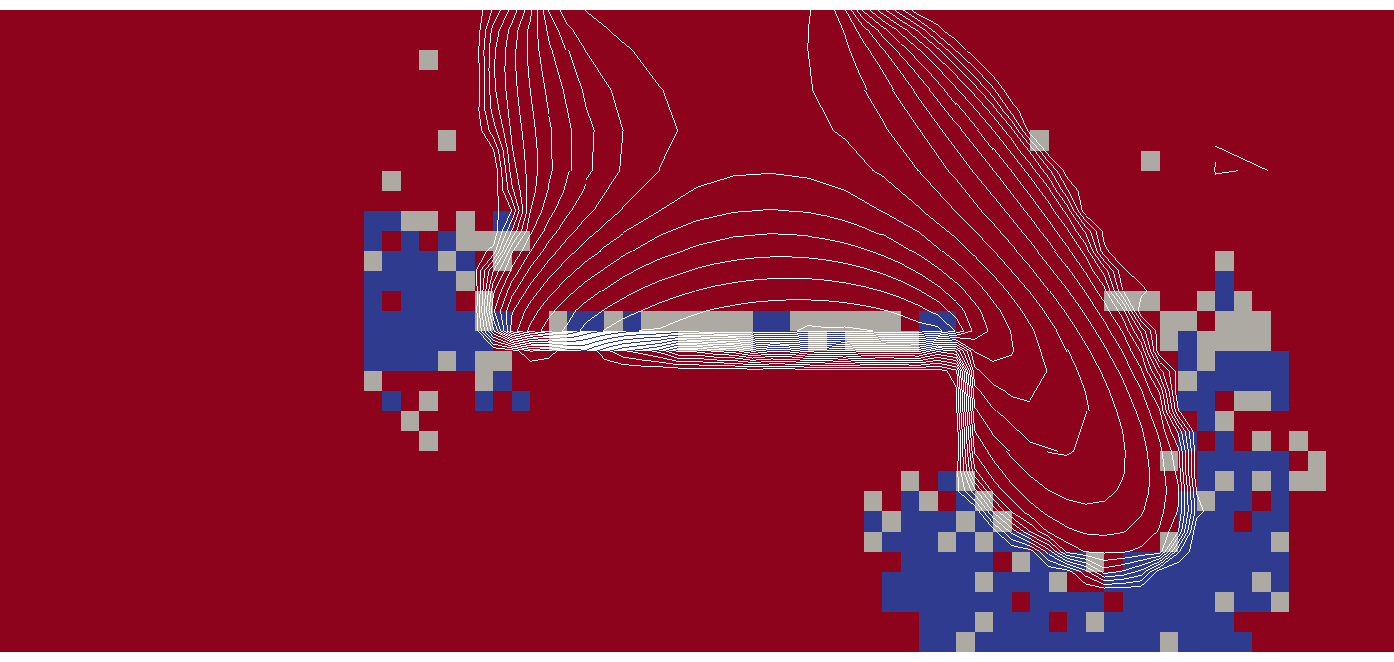}

  \includegraphics[width=0.32\textwidth]{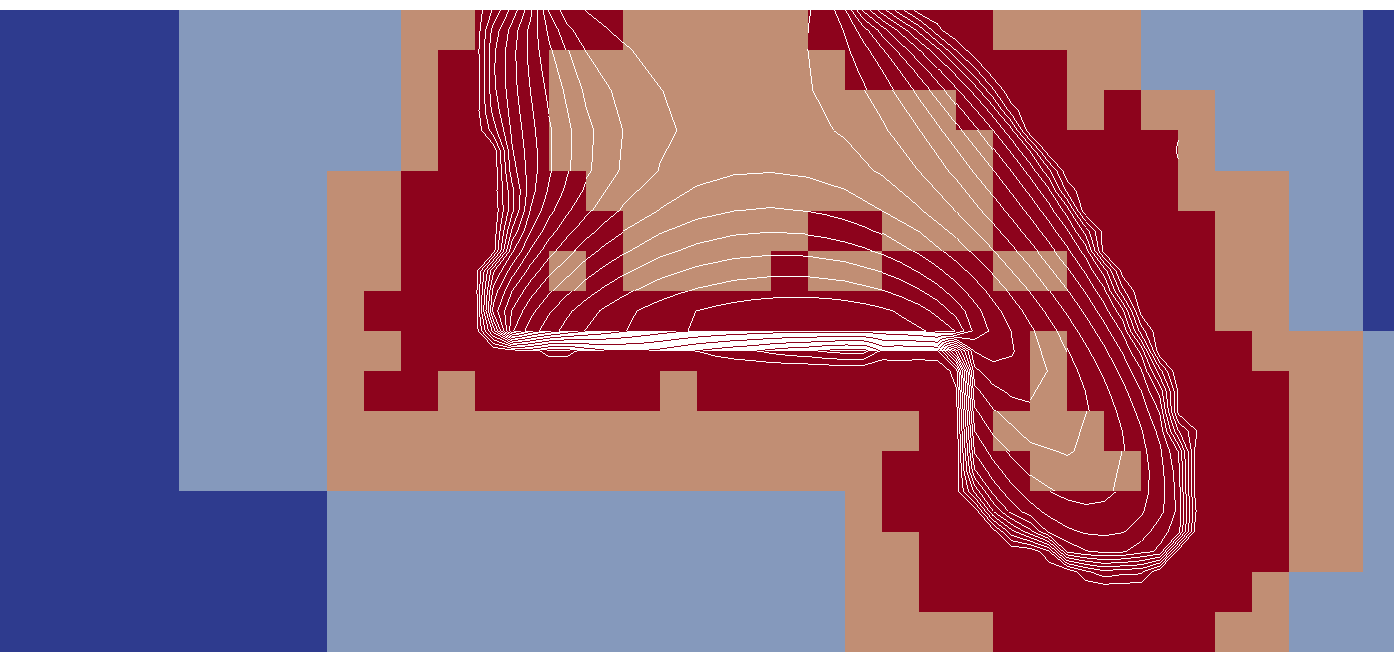}
  \includegraphics[width=0.32\textwidth]{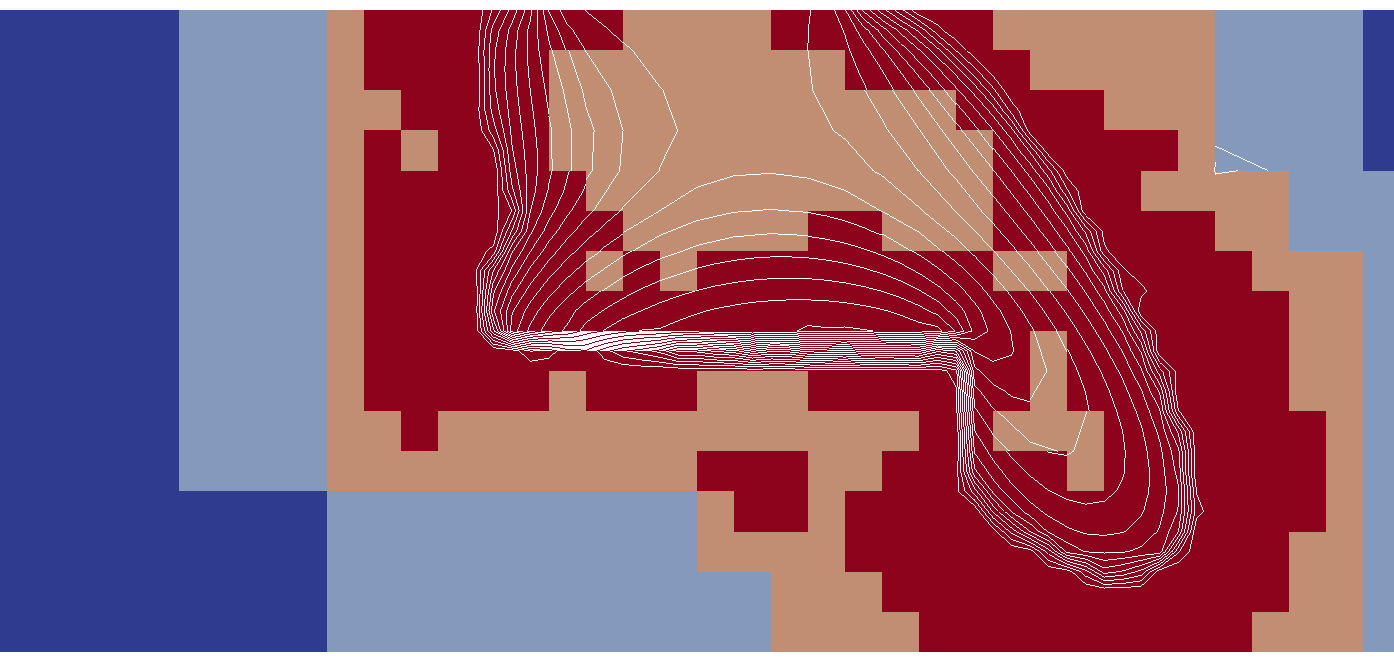}
  \includegraphics[width=0.32\textwidth]{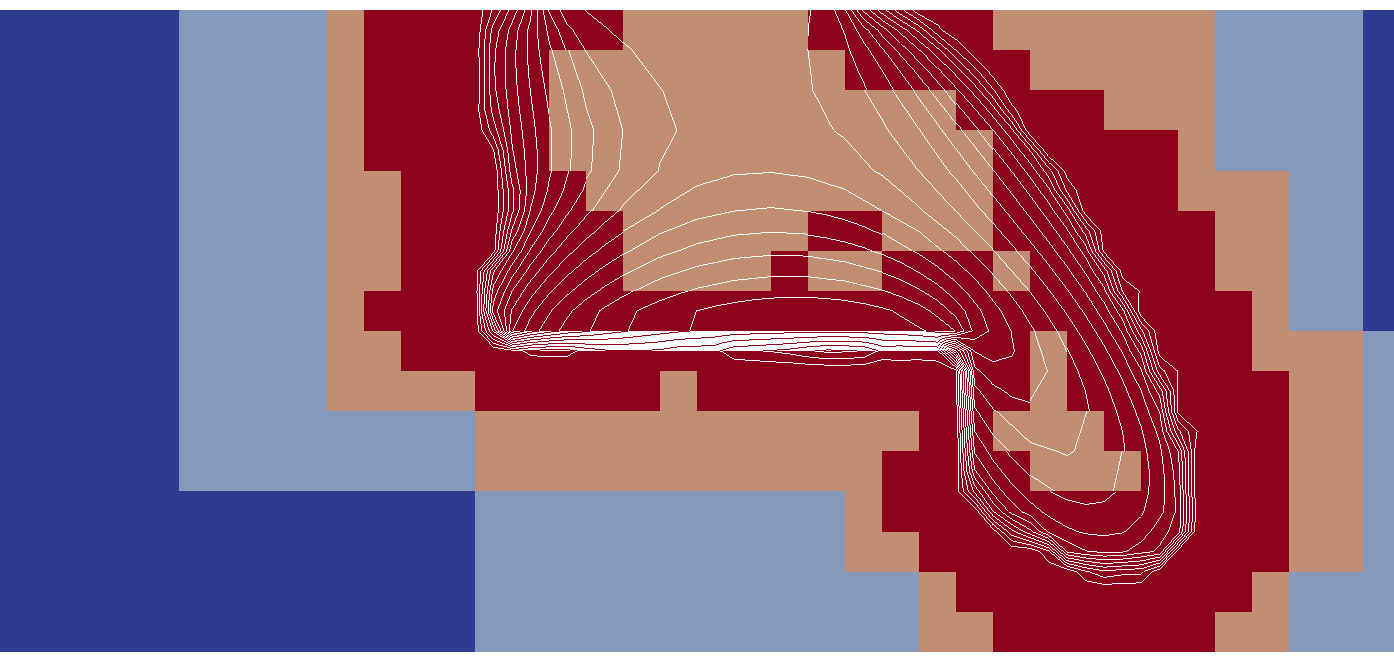}
  \caption{Comparison of different approaches for choosing the local
  polynomial degree.
  The top row shows the distribution of the polynomial
  order (left: original approach, right: modified indicator).
  Red color refers to $r=3$ and blue refers to $r=1$.
  The bottom row shows the grid level used for the three simulation.
  Red refers to $l=3$ and blue refers to $l=0$.
  Left to right: original approach, modified indicator, with uniform
  polynomial degree of three. White lines show 20 contour levels between
  $s_n=0$ and $s_n=0.55$.}
% sat33_implicit_dt5_nop.png
% sat33_implicit_dt5.png
% sat33_implicit_dt5_smoothness.png
\label{fig:padaptive_order_level}
\end{figure}

\begin{figure}[!ht]
  \centering
  \includegraphics[width=0.475\textwidth]{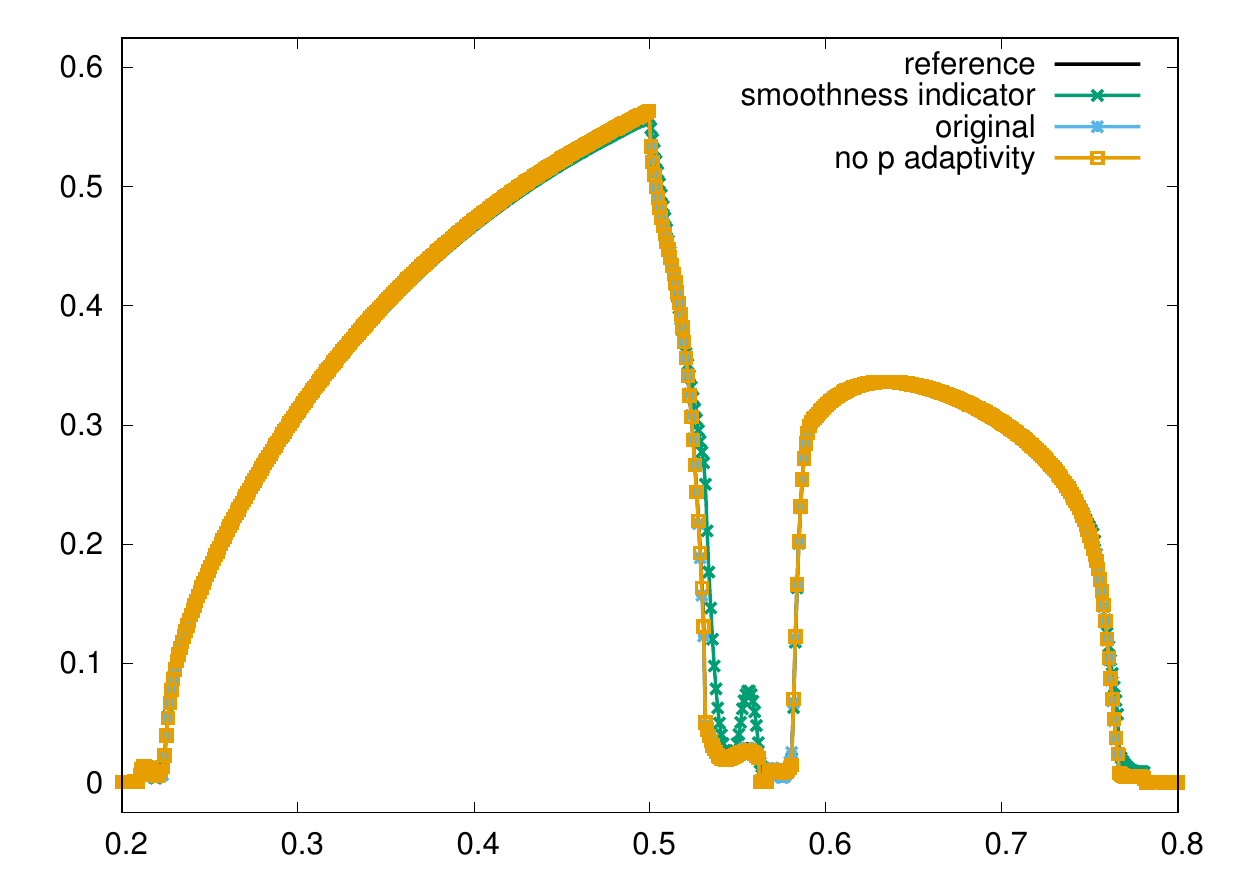}
  \includegraphics[width=0.475\textwidth]{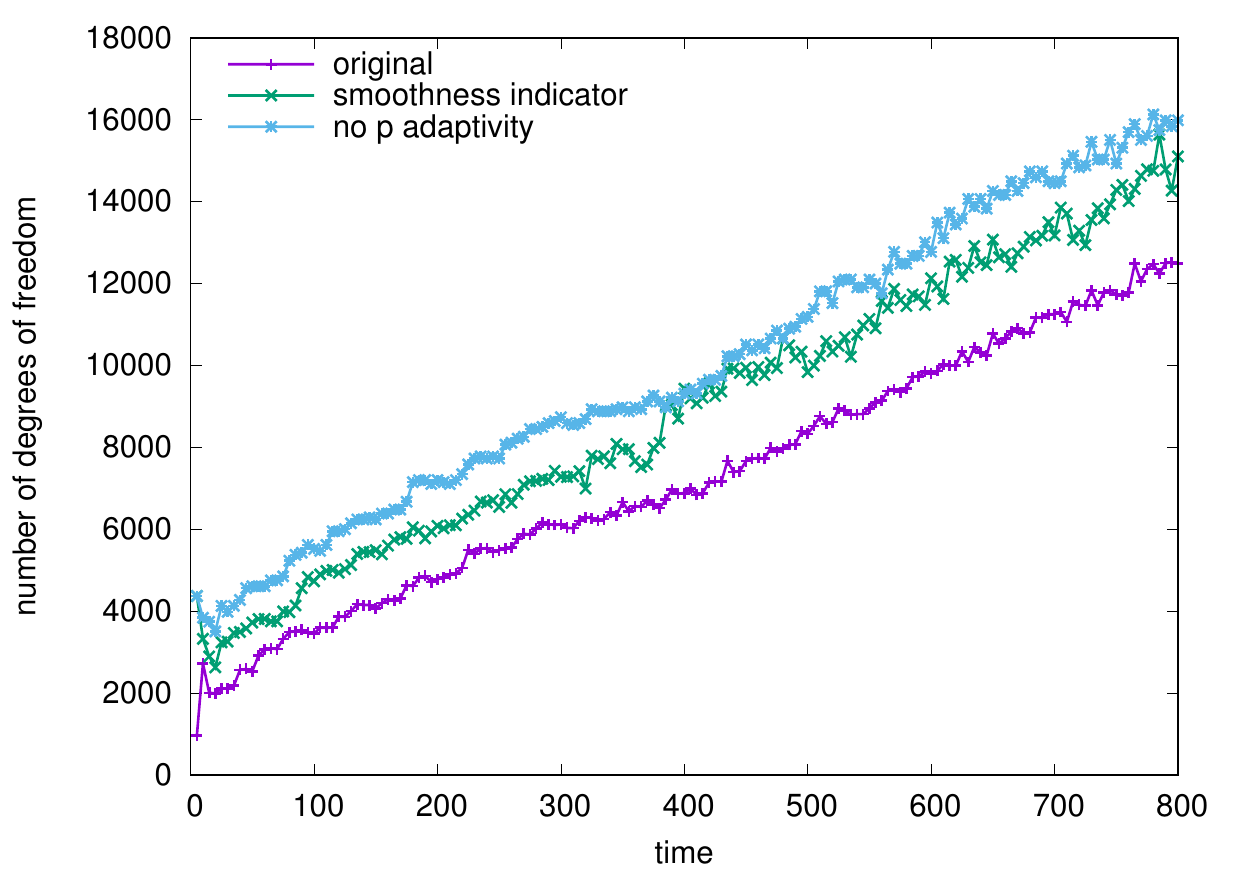}
  \caption{Comparison of different approaches for choosing the local
  polynomial degree. Shown $s_n$ over the line in equation \eqref{eq:plotline}. }
  \label{fig:padaptive_line_dofs}
\end{figure}

Figure~\ref{fig:padaptive_order_level} shows the distribution of the
polynomial order for the two p-adaptive approaches. The local grid
adaptivity is of course also influenced by the choice of indicator for the
polynomial degree because the values of the residuals change. This can
also be seen in Figure~\ref{fig:padaptive_order_level}. As expected for the
approach discussed in Section~\ref{sec:adaptivity} the polynomial order is
reduced to the smallest admissible value ($r=1$) in the regions where $s_n$
is constant.
The approach based on the smoothness indicator described in this
section clearly leads to a reduction in the polynomial order at the
interface of the plume. This is expected since here the solution can be
said to have lower regularity. This demonstrates that the indicator works as
expected. On the left in Figure~\ref{fig:padaptive_line_dofs} the solution for the different approaches along the same line given in equation \eqref{eq:plotline} is shown.
While the solution for the original adaptive method and the solution
without adaptivity are indistinguishable, the solution with the smoothness
indicator given above shows clearly an increase in numerical diffusion at
the interfaces - especially at the entry point to the lens which also leads
to an increase within the lens.
The number of degrees of freedom during the course of the simulation
depends on the number of elements and the local distribution of the
polynomial degree used. The number of elements increases in time and so
does the number of degrees of freedom. The right plot in
Figure~\ref{fig:padaptive_line_dofs} shows the number of degrees of freedom
as a function of time. Clearly the method with maximal polynomial degree on
all elements requires the most degrees of freedom. Using the original
indicator of p-adaptivity reduces the number of degrees of freedom to about
$66\%$ at the beginning of the simulation and still to
$75\%$ at the final time. The smoothness indicator given in this section only leads to a
reduction of $20\%$ at the beginning of the simulation and by only $6\%$ at
the final time.

Overall the indicator described in Section~\ref{sec:adaptivity} does seem to lead to a better distribution of the polynomial degree with negligible influence of the actual solution. As can be seen from Figure~\ref{fig:padaptive_order_level}, there is only little reduction of the order in the actual plume and the intermediate order $p=2$ is hardly used anywhere in the domain. Both these observations indicate that further research into p-adaptivity for this type of problem is required.

\subsection{Isotropic Flow over a weak Lens}
In the final section we just study a second test case. The setup is the
same as in the previous example but the permeability tensors is isotropic
and the lens is weaker:
\begin{align*}
  \mathbb{K}_{\Omega  \backslash \Omega_{lens}} &=
  \begin{pmatrix}\label{}
  10^{-10}&  0 \\
  %& &\\
  0 & 10^{-10}
  \end{pmatrix}m^2~, &&
  \mathbb{K}_{ \Omega_{lens}} &=
  \begin{pmatrix}\label{}
  10^{-12}&  0\\
  0 & 10^{-12}
  \end{pmatrix}m^2~.
\end{align*}
On the Python side the problem class the definition of $K$ has to be
modified accordingly as shown in \ref{sec:codeModifications_iwl}.

\begin{figure}
  \includegraphics[width=0.24\textwidth]{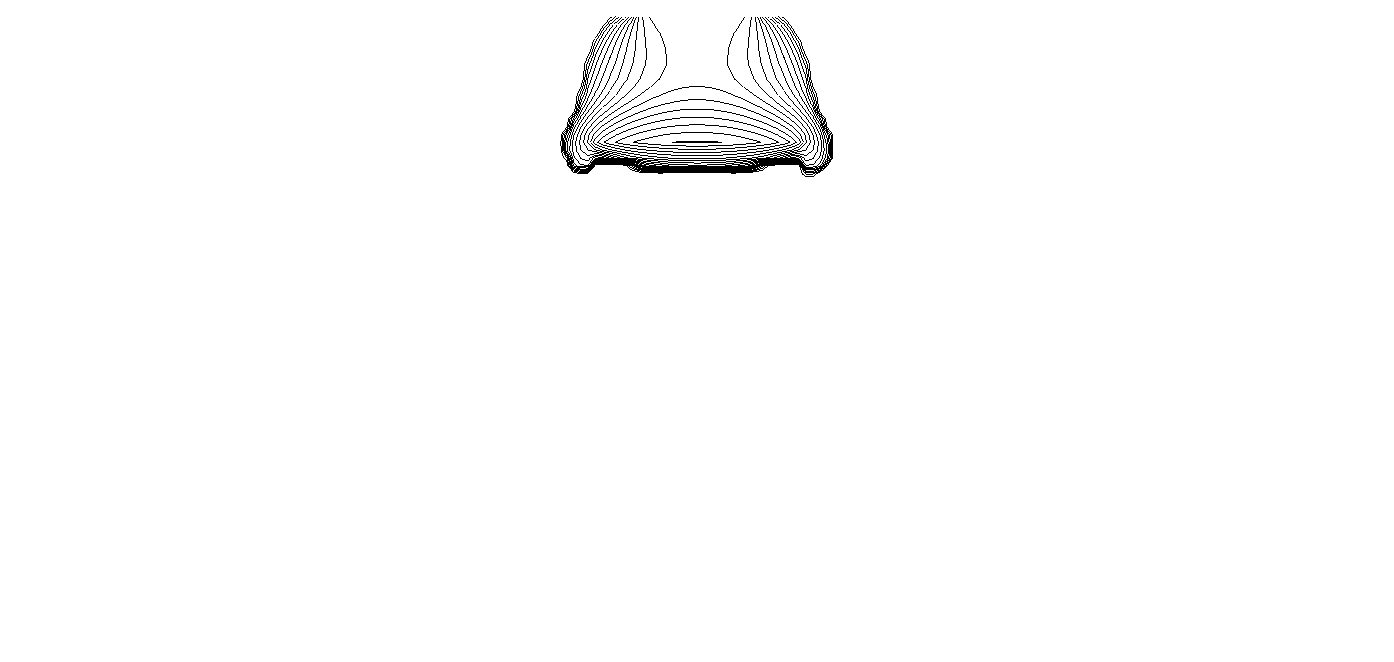}
  \includegraphics[width=0.24\textwidth]{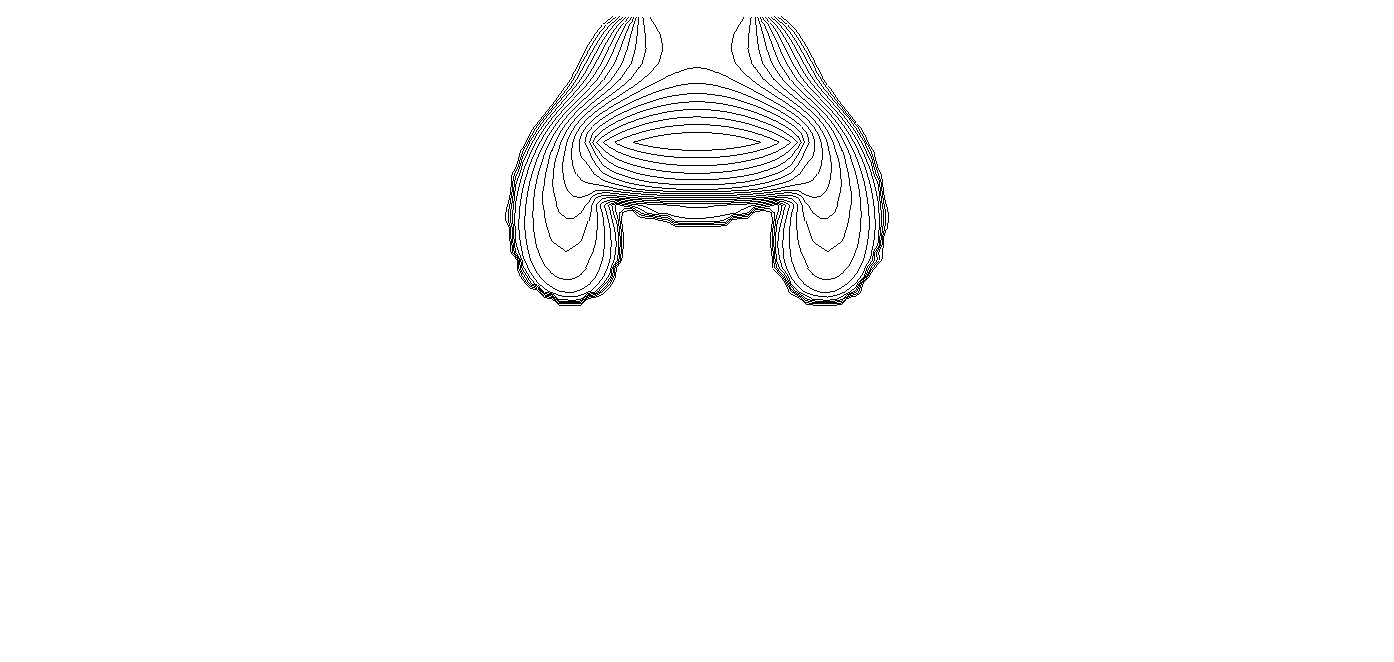}
  \includegraphics[width=0.24\textwidth]{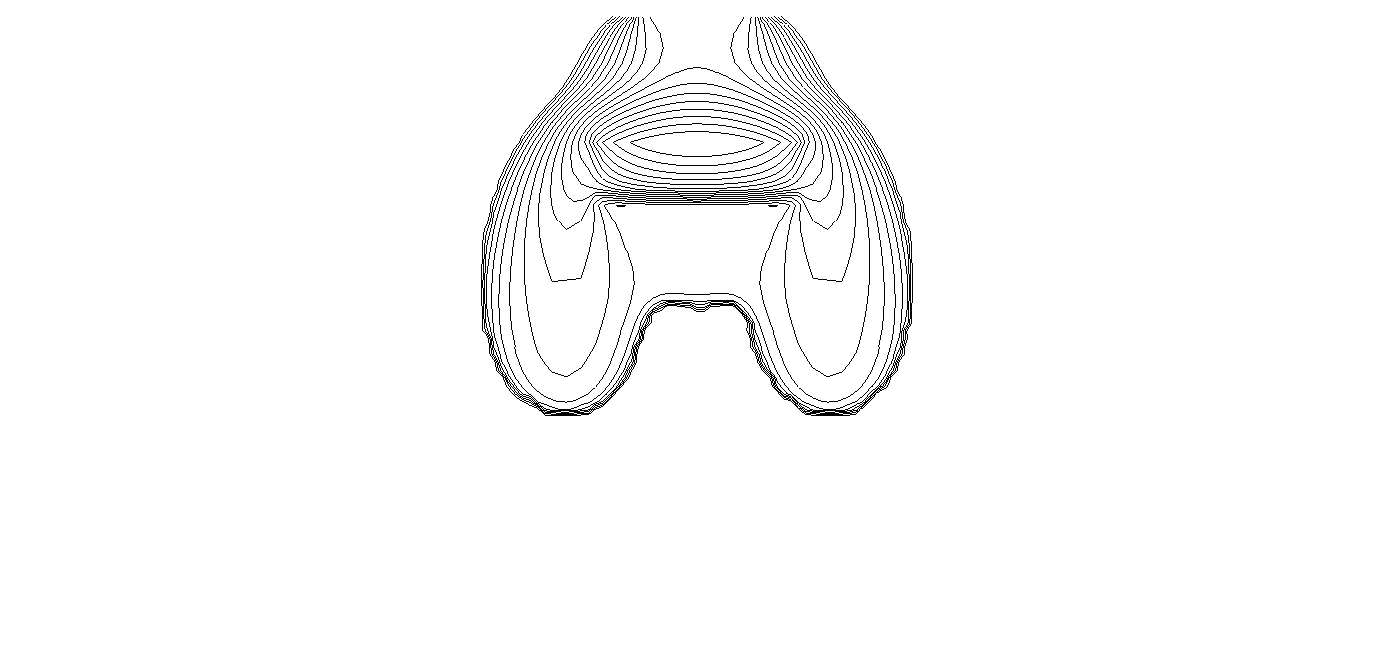}
  \includegraphics[width=0.24\textwidth]{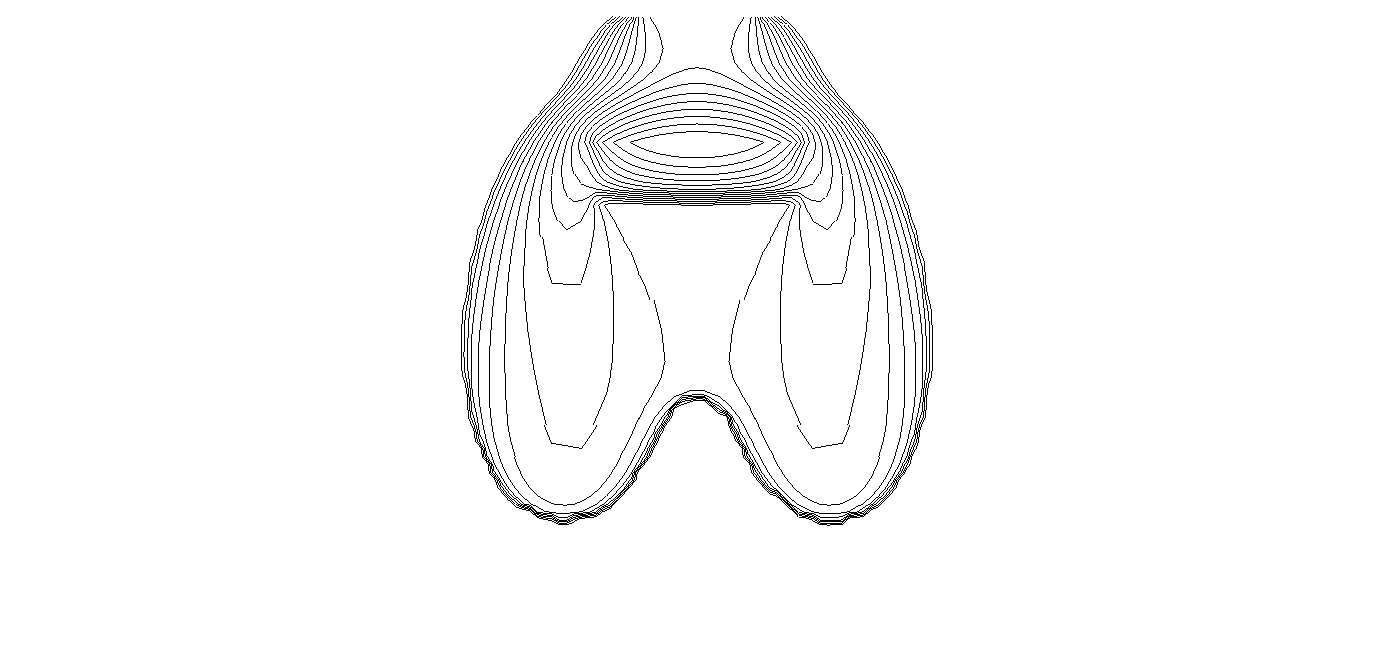}

  \includegraphics[width=0.24\textwidth]{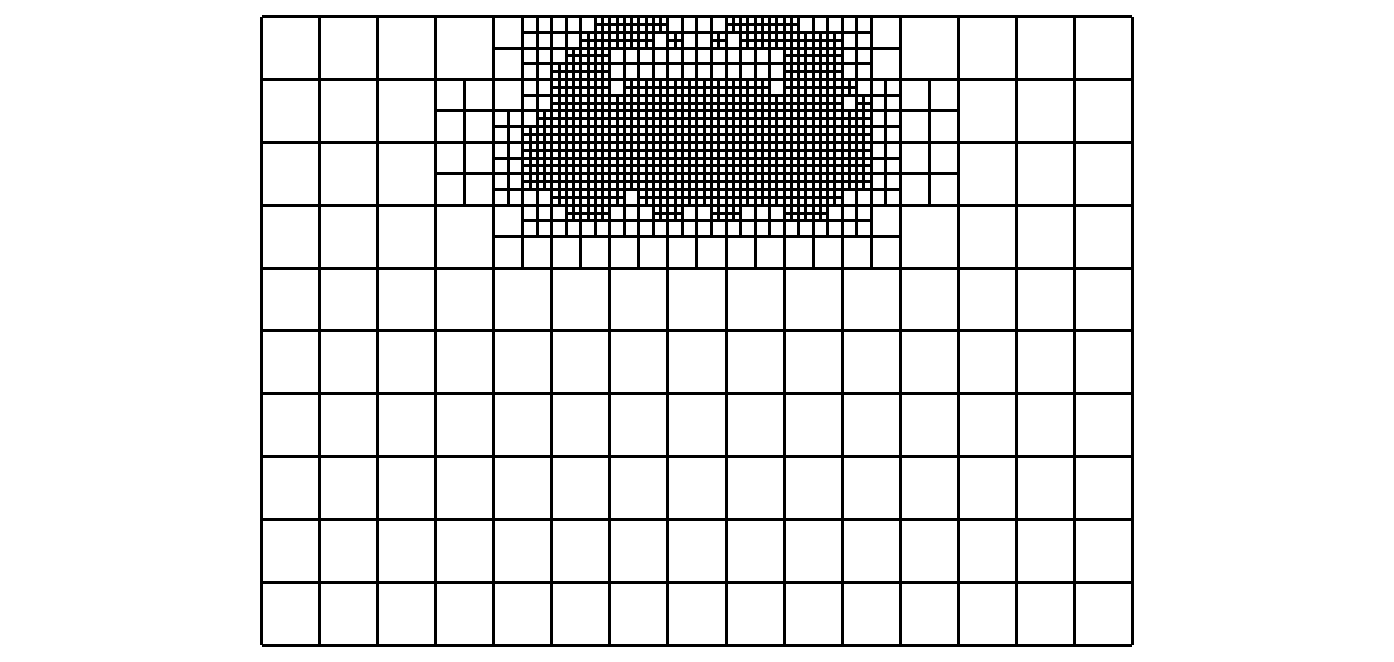}
  \includegraphics[width=0.24\textwidth]{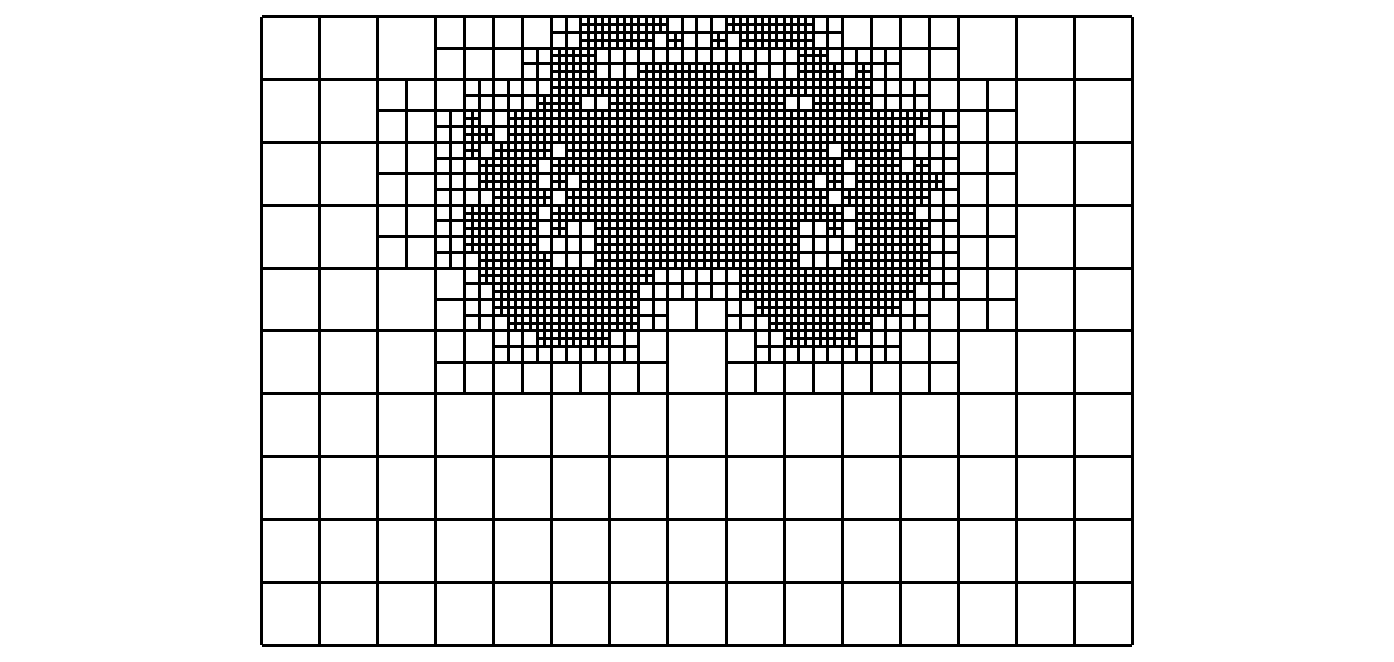}
  \includegraphics[width=0.24\textwidth]{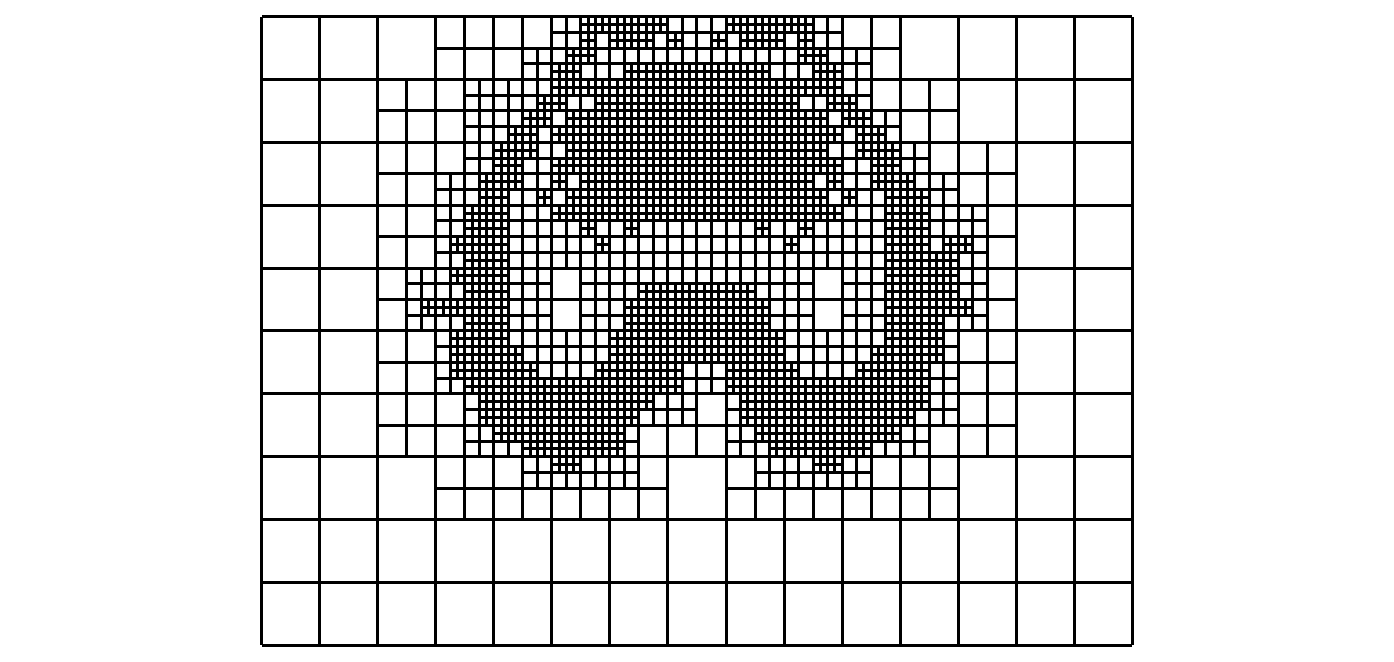}
  \includegraphics[width=0.24\textwidth]{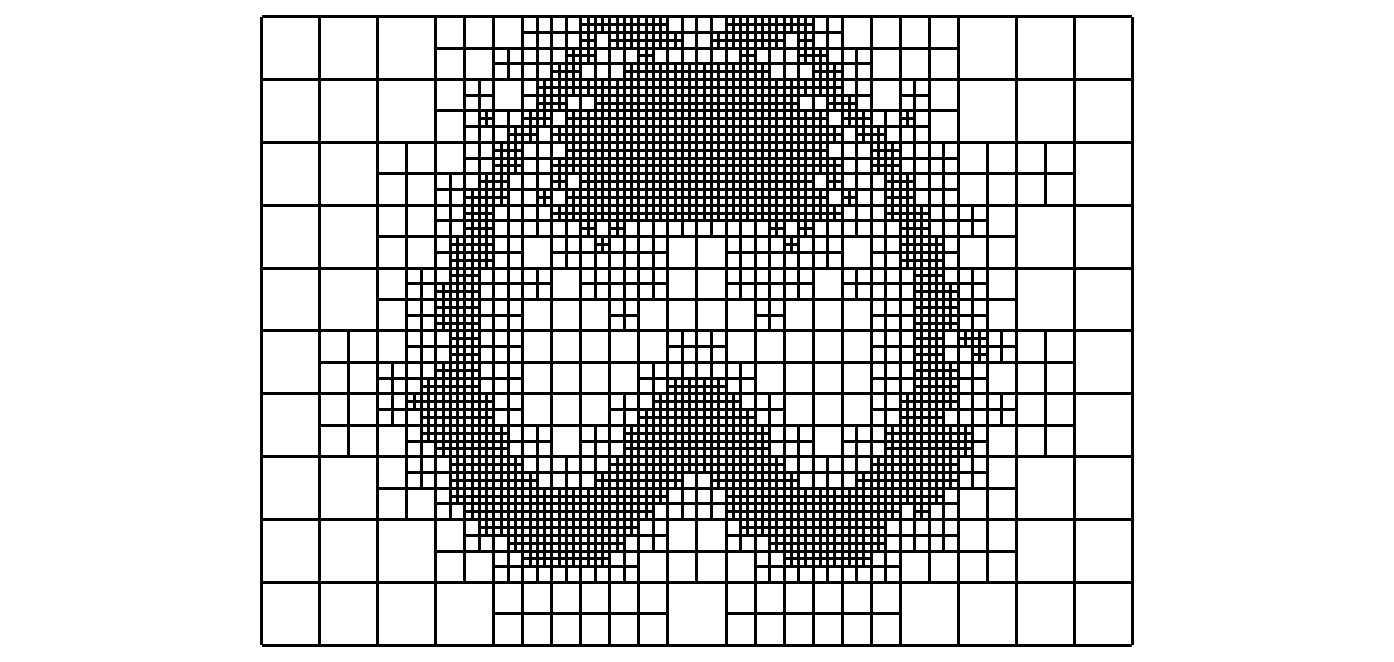}
  \caption{Evolution of the non wetting saturation $s_n$ for the isotropic
           weak lens setting
           at times $t=800,1600,2400$, and $t=3200$ (top) and the
           corresponding adaptive grid structure (bottom).}
  \label{fig:flowevolution_iwl}
\end{figure}

Snapshots of the evolution of the resulting flow are shown in
Figure~\ref{fig:flowevolution_iwl}. The symmetry is clearly visible both in
the solution and in the grid refinement. The grid is locally refined at the
interface and around the lens once the flow reaches that point. Dune to the
weaker lens flow also passes through the lens. Overall our tests indicate
that the conclusions obtained from our previous tests also apply to this
setting.

% provide details of the constants, tolerances, stopping criteria etc. used
% Then show results for the anisotropic lens

\section{Conclusion}
\label{sec:conclusion}

We have presented a framework that allows to study hp-adaptive schemes for
two-phase flow in porous media. The presented approach allows to easily study
different adaptive strategies and time stepping algorithms. 
The change from one algorithm to another is easily implemented. All python based
implementation is in the end forwarded to C++ based implementations to ensure
performance of the applications.

Furthermore, the prototypes build on very mature implementations from the
\texttt{Dune} community to allow for a short transition from prototypes to
production codes. Parallelization and extension to 3d are straightforward with
the prototype presented.

We focused in this paper mostly on the stability of different approaches
for evolving the solution from one time step to the next. Our results
indicate that IMPES type splitting schemes can be used when smaller time
step are acceptable while for larger time steps schemes solving the fully
coupled equations should be used. The simple fixed point iteration seems to
work quite well and seems quite stable with respect to the algorithmic
approaches and models tested. It turns out that the method does show
convergence issues when too small values for the stopping tolerance are used.
The Newton method has more difficulty with convergence for large time
steps. This suggests a combined approach where first the iterative scheme
is used to reach a reasonable starting point for the Newton solver. We
tested this approach and could obtain reasonable results up to time steps
of $\tau=25$ effectively tripling the maximum time steps achievable when
using only the Newton scheme. In future work we will focus more on this
approach.

In addition, we will investigate 3d examples and the possible extension to polyhedral
cells which are widely used in industrial applications. Preliminary work has been carried out, 
for example, in \cite{reconpoly:17}. The deployment of higher order adaptive schemes is an essential tool for
capturing reactive flows for applications such as polymer injections for
improved oil recovery or CO$_2$ sequestration. Here, improved numerical algorithms
help to reduce uncertainty for predictions and thus ultimately improve 
decision making capabilities for involved stakeholders.

%\todo[inline]{Talk about possible extensions including work on polyhedral grids (cite Robert + Martin +Anna), Compressible multiphase, 3d examples, multicomponents} 

\section*{Acknowledgements}

Birane Kane acknowledges the Cluster of Excellence in Simulation Technology (SimTech) at
the University of Stuttgart for financial support.
Robert Kl\"ofkorn acknowledges the Research Council of Norway and the 
industry partners, ConocoPhillips Skandinavia AS, Aker BP ASA, 
Eni Norge AS, Maersk Oil; a company by Total, 
Statoil Petroleum AS, Neptune Energy Norge AS, Lundin Norway AS, 
Halliburton AS, Schlumberger Norge AS, Wintershall Norge AS, and DEA Norge AS, 
of The National IOR Centre of Norway for support.

%All authors would like to thank the reviewers for helpful comments to improve
%this work.

%\section*{References}
\refsection

%\bibliographystyle{elsarticle-num}
%\bibliography{bibliography}

\appendix
\section{Reproducing the Results in a Docker Container}
\label{sec:docker}

To easily reproduce the results of this paper we provide a Docker \cite{boettinger:15}
image containing the presented code in a Jupyter notebook \cite{kluyver:16} and all
necessary software to run it.

Once Docker is installed, the following shell command will start the Jupyter server
within a Docker container:
\begin{lstlisting}[language=bash]
docker run --rm -v dune:/dune -p 127.0.0.1:8888:8888 registry.dune-project.org/dune-fem/twophaseflow:latest
\end{lstlisting}
Notice that all user data will be put into and kept in the Docker volume named
\texttt{dune} for later use.
This volume should not exist prior to the first run of the above command.

Open your favorite web browser and connect to \texttt{127.0.0.1:8888} and log in;
the password is \texttt{dune}.
The notebook \texttt{twophaseflow} contains the code used to obtain the results in
this paper.

\section{Main Code Structure}
\label{sec:code}

In this section we show parts of the python script used in the simulation.
The snippets are not self contained but should provide enough information to
understand the overall structure. The full code which can be used to
produce the simulations in Section~\ref{sec:timestability} is available as
a jupyter notebook (see Appendix~\ref{sec:docker}) for details. In this section we
describe parts of the code following the overall structure of
Sections~\ref{sec:model} and \ref{sec:discretization}.

\subsection{Model A: Wetting-phase-pressure/nonwetting-phase-saturation formulation}
The model description is decomposed into two parts: this first part
consists of a \emph{problem} class containing a static function for the
pressure law $p_c$, the permeability tensor $K$, boundary, and initial
data, and the further constants needed to fully describe the problem:
\begin{python}
  class AnisotropicLens:
      dimWorld = 2
      x     = SpatialCoordinate(triangle)

      g     = [0,]*dimWorld ; g[dimWorld-1] = -9.810 # [m/s^2]
      g     = as_vector(g)
      r_w   = 1000.  # [Kg/m^3]
      mu_w  = 1.e-3  # [Kg/m s]
      r_n   = 1460.  # [Kg/m^3]
      mu_n  = 9.e-4  # [Kg/m s]

      lensDomain = conditional(abs(x[1]-0.49)<0.03,1.,0.)*\
                   conditional(abs(x[0]-0.45)<0.11,1.,0.)

      lens = lambda a,b: a*lensDomain + b*(1.-lensDomain)

      Kdiag = lens(6.64*1e-14, 1e-10) # [m^2]
      Koff  = lens(0,-5e-11)          # [m^2]
      K     = as_matrix( [[Kdiag,Koff],[Koff,Kdiag]] )

      Phi   = lens(0.39, 0.40)             # [-]
      s_wr  = lens(0.10, 0.12)             # [-]
      s_nr  = lens(0.00, 0.00)             # [-]
      theta = lens(2.00, 2.70)             # [-]
      pd    = lens(5000., 755.)            # [Pa]

      #### initial conditions
      p_w0 = (0.65-x[1])*9810.       # hydrostatic pressure
      s_n0 = 0                       # fully saturated
      # boundary conditions
      inflow = conditional(abs(x[0]-0.45)<0.06,1.,0.)*\
               conditional(abs(x[1]-0.65)<1e-8,1.,0.)
      J_n  = -5.137*1e-5
      J_w  = 1e-20        # ufl bug?
      dirichlet = conditional(abs(x[0])<1e-8,1.,0.) +\
                  conditional(abs(x[0]-0.9)<1e-8,1.,0.)
      p_wD = p_w0
      s_nD = s_n0

      q_n  = 0
      q_w  = 0

      p_c = brooksCorey
\end{python}
The Brooks-Corey pressure law is given by a function taking a
\emph{problem} class as first argument and the value non wetting phase
$s_n$:
\begin{python}
  def brooksCorey(P,s_n):
      s_w = 1-s_n
      s_we = (s_w-P.s_wr)/(1.-P.s_wr-P.s_nr)
      s_ne = (s_n-P.s_nr)/(1.-P.s_wr-P.s_nr)
      cutOff = lambda a: min_value(max_value(a,0.00001),0.99999)
      if P.useCutOff:
          s_we = cutOff(s_we)
          s_ne = cutOff(s_ne)
      kr_w = s_we**((2.+3.*P.theta)/P.theta)
      kr_n = s_ne**2*(1.-s_we**((2.+P.theta)/P.theta))
      p_c  = P.pd*s_we**(-1./P.theta)
      dp_c = P.pd * (-1./P.theta) * s_we**(-1./P.theta-1.) * (-1./(1.-P.s_wr-P.s_nr))
      l_n  = kr_n / P.mu_n
      l_w  = kr_w / P.mu_w
      return p_c,dp_c,l_n,l_w
\end{python}

The actual PDE description requires three vector valued coefficient functions
one for the solution on the new time level (\pyth{u}), one for the solution on the previous
time level (\pyth{solution_old}), and one for the intermediate state $\bar{s}$
used in the iterative approaches (\pyth{intermediate}). The vector valued
test function is \pyth{v}. Furthermore, $\tau,\beta$ are constants used
used for the time step size, the penalty factor, respectively. These can be
set dynamically during the simulation:
\begin{python}
  s_n  = u[1]
  s_w  = 1.-s_n
  si_n = intermediate[1]
  si_w = 1.-si_n

  p_c,dp_c,l_n,l_w = P.p_c(s_n=si_n)

  p_w  = u[0]
  p_n  = p_w + p_c
  gradp_n = grad(p_w) + dp_c * grad(s_n)

  velocity_n = P.K*(gradp_n-P.r_n*P.g)
  velocity_w = P.K*(grad(p_w)-P.r_w*P.g)

  #### bulk equations
  dbulk_p  = P.K*( (l_n+l_w)*grad(p_w) + l_n*dp_c*grad(s_n) )
  dbulk_p += -P.K*( (P.r_n*l_n+P.r_w*l_w)*P.g )
  bulk_p   = P.q_w+P.q_n
  dbulk_s  = P.K*l_n*dp_c*grad(s_n)
  dbulk_s += P.K*l_n*(grad(p_w)-P.r_n*P.g)
  bulk_s   = P.q_n
\end{python}

\subsection{Space Discretization}
Given the expressions defined previously the bulk integrals for the bilinear forms can now be easily defined
(compare Section~\ref{sec:discretization}):
\begin{python}
  form_p = ( inner(dbulk_p,grad(v[0])) - bulk_p*v[0] ) * dx
  form_s = ( inner(dbulk_s,grad(v[1])) - bulk_s*v[1] ) * dx
  form_p += J_p * v[0] * P.inflow * ds
  form_s += J_s * v[1] * P.inflow * ds
\end{python}
Next we describe the skeleton terms required for the DG formulation.
We use some geometric terms defined for the skeleton of the grid and also
the weighted average:
\begin{python}
  def sMax(a): return max_value(a('+'), a('-'))
  def sMin(a): return min_value(a('+'), a('-'))
  n         = FacetNormal(cell)
  hT        = MaxCellEdgeLength(cell)
  he        = avg( CellVolume(cell) ) / FacetArea(cell)
  heBnd     = CellVolume(cell) / FacetArea(cell)
  k         = dot(P.K*n,n)
  def wavg(z): return (k('-')*z('+')+k('+')*z('-'))/(k('+')+k('-'))
\end{python}
As shown in Section~\ref{sec:discretization} it is straightforward to construct the
required penalty and consistency terms
\begin{python}
  ## penalty
  form_p  = penalty_p[0]/he * jump(u[0])*jump(v[0]) * dS
  form_s  = penalty_s[0]/he * jump(u[1])*jump(v[1]) * dS
  ## consistency
  form_p -= inner(wavg(dBulk_p),n('+')) * jump(v[0]) * dS
  form_s -= inner(wavg(dBulk_s),n('+')) * jump(v[1]) * dS

  ##### dirichlet conditions
  ## penalty
  form_p += penalty_p[1]/heBnd * (u[0]-p_D) * v[0] * P.dirichlet * ds
  form_s += penalty_s[1]/heBnd * (u[1]-s_D) * v[1] * P.dirichlet * ds
  ## consistency
  form_p -= inner(dBulk_p,n) * v[0] * P.dirichlet * ds
  form_s -= inner(dBulk_s,n) * v[1] * P.dirichlet * ds
\end{python}
The factors for the penalty terms for the DG discretization depend on
the model and are given by
\begin{python}
  lambdaMax = k('+')*k('-')/avg(k) # P.K[0][0] + abs(P.K[0][1])  # assuming 2d and K=[[a,b],[b,a]]
  p_c0,dp_c0,l_n0,l_w0 = P.p_c(0.5) # is not the maximm (increases for s_n->1)
  penalty_p     = beta*lambdaMax*sMax(l_n0+l_w0)
  penalty_s     = beta*lambdaMax*sMax(l_n0*dp_c0)
  penalty_bnd_p = beta*k*(l_n0+l_w0)
  penalty_bnd_s = beta*k*(l_n0*dp_c0)
\end{python}

\subsection{Time stepping}
The final bilinear forms used to carry out the time stepping depend on the
actual schemes used. We first need to distinguish between the three
schemes \emph{linear,implicit,iterative} that are based on the full coupled
system and the two schemes \emph{impes,iterative-impes} which are based on
a decoupling of the pressure and saturation equation. In the first case the
final bilinear form is simply
\begin{python}
  form = form_s + form_p
\end{python}
while in the second case we define a pair of scalar forms:
\begin{python}
  uflSpace1 = Space((problem.dimWorld,problem.dimWorld),1)
  u1        = TrialFunction(uflSpace1)
  v1        = TestFunction(uflSpace1)
  form_p = replace(form_p, { u:as_vector([u1[0],intermediate.s[0]]),
                             v:as_vector([v1[0],0.]) } )
  form_s = replace(form_s, { u:as_vector([solution[0],u1[0]]),
                             intermediate:as_vector([solution[0],
                                                     intermediate[1]]),
                             v:as_vector([0.,v1[0]]) } )
  form = [form_p,form_s]
\end{python}
Finally we need to fix $\bar{s}$ i.e. \pyth{intermediate} according to the
scheme used. In the case of the \emph{implicit} scheme we have
\pyth{intermediate=u}, for \emph{linear} and \emph{impes}
\pyth{intermediate=solution_old}, while for the other two schemes
\pyth{intermediate} is an independent function used during the iteration.

The following code demonstrates how the evolution of the solution from
$t^i$ to $t^{i+1}$ is carried out:
\begin{python}
  while True:
      intermediate.assign(solution)
      scheme.solve(target=solution)
      limit( solution )
      if errorMeasure(solution,solution-intermediate)
          break
\end{python}
where the stopping criteria is given by
\begin{python}
  def errorMeasure(w,dw):
      rel = integrate(grid, [w[1]**2,dw[1]**2], 5)
      tol = self.tolerance * math.sqrt(rel[0])
      rdiff = math.sqrt(rel[1])
      return rdiff < tol
\end{python}
The implementation of the \emph{iterative-impes} method looks almost the same
\begin{python}
  while n<self.maxIterations:
      intermediate.assign(solution)
      limit( iterate )
      scheme[0].solve(target=solution.p)
      scheme[1].solve(target=solution.s)
      limit( solution )
      n += 1
      if error(solution,solution-intermediate):
          break
\end{python}

\subsection{Stabilization}
Note how we apply the limiting operator directly after the next iterate has
been computed. The stabilization projection operator is available as \pyth{limit( solution )}.

\subsection{Adaptivity}
The estimator is given as a form taking vector valued solution
\pyth{u} with a scalar test function \pyth{v0}. This will later be used to
generate an operator taking the solution and mapping into a piece wise
constant scalar space with the value $\eta_E$ on each element:
\begin{python}
  uflSpace0 = Space((P.dimWorld,P.dimWorld),1)    # space for indicator (could use dimRange=3)
  v0        = TestFunction(uflSpace0)

  Rvol = P.Phi*(u[1]-solution_old[1])/tau - div(dBulk_s) - bulk_s
  estimator = hT**2 * Rvol**2 * v0[0] * dx +\
          he * inner(jump(dBulk_s), n('+'))**2 * avg(v0[0]) * dS +\
          heBnd * (J + inner(dBulk_s,n))**2 * v0[0] * P.inflow * ds +\
          penalty_s[0]**2/he * jump(u[1])**2 * avg(v0[0]) * dS +\
          penalty_s[1]**2/heBnd * (s_D - u[1])**2 * v0[0] * P.dirichlet * ds
\end{python}
and since we want to use the estimator for the fully coupled implicit
problem independent of the actual time stepping approach used, we add
\begin{python}
  estimator = replace(estimator, {intermediate:u})
\end{python}
The actual grid adaptivity is then carried out by calling:
\begin{python}
  estimator(solution, estimate)
  hgrid.mark(markh)
  fem.adapt(hgrid,[solution])
\end{python}
where the marking function is
\begin{python}
hTol = 1e-16 # initial value, later tTol * dt / gridSize
def markh(element):
    estimateLocal = estimate.localFunction(element)
    r = estimateLocal.evaluate(element.geometry.referenceElement.center)
    eta = sum(r)
    if eta > hTol and element.level < maxLevel:
        return Marker.refine
    elif eta < 0.01*hTol:
        return Marker.coarsen
    else:
        return Marker.keep
\end{python}
compare Algorithm~\ref{alg:h-adapt}.

Finally the p-adaptivity requires calling
\begin{python}
  estimator(solution, estimate)
  # project solution to space with p-1
  orderreduce(solution,sol_pm1)
  # compute estimator for p-1 space
  estimator(sol_pm1, estimate_pm1)
  # compute smoothness indicator and modify polynomial order
  fem.spaceAdapt(space, markp, [solution])
\end{python}
where the marking function \pyth{markp} is given by
\begin{python}
  def markp(element):
      r      = estimate.localFunction(element).evaluate(center)[0]
      r_p1   = estimate_pm1.localFunction(element).evaluate(center)[0]
      eta = abs(r-r_p1)
      polorder = spc.localOrder(element)
      if eta < pTol:
          return polorder-1 if polorder > 1 else polorder
      elif eta > 100.*pTol:
          return polorder+1 if polorder < maxOrder else polorder
      else:
          return polorder
\end{python}
compare Algorithm~\ref{alg:p-adapt}.

%%%%%%%%%%%%%%%%%%%%%%%%%%%%%%%%%%%%%%%%%%%%%%%%%%%%%%%%%%%%%%%%%%%%%%%%%%%%%%%

\section{Code Modifications}
\label{sec:codeModifications}

\subsection{Cut off stabilization}
\label{sec:codeModifications_cufoff}

The \emph{cut off stabilization} can be easily implemented with a minor
change to the function defining the capillary pressure:

\begin{python}
def brooksCorey(P,s_n):
    # cut all values of s below 1e-5 and above 0.99999
    s_w = 1-s_n
    cutOff = lambda a: min_value(max_value(a,0.00001),0.99999)
    s_we = cutOff( (s_w-P.s_wr)/(1.-P.s_wr-P.s_nr) )
    s_ne = cutOff( (s_n-P.s_nr)/(1.-P.s_wr-P.s_nr) )
    kr_w = s_we**((2.+3.*P.theta)/P.theta)
    kr_n = s_ne**2*(1.-s_we**((2.+P.theta)/P.theta))
    p_c  = P.pd*s_we**(-1./P.theta)
    dp_c = P.pd * (-1./P.theta) * s_we**(-1./P.theta-1.) * (-1./(1.-P.s_wr-P.s_nr))
    l_n  = kr_n / P.mu_n
    l_w  = kr_w / P.mu_w
    return p_c,dp_c,l_n,l_w
\end{python}

\subsection{Different model: Model B}
\label{sec:codeModifications_modelB}

Changing the formulation of the two phase flow model requires redefining
the terms for the bulk integrals and the penalty factor for the DG
stabilization. The adaptation indicators and other DG terms do not need to
be touched:
\begin{python}
  s_n    = u[1]
  p_avg  = u[0]
  p_c,dp_c,l_n,l_w = P.p_c(intermediate[1])

  dBulk_p  =  P.K*( (l_n+l_w)*grad(p_avg) + 0.5*(l_n-l_w)*dp_c*grad(s_n) )
  dBulk_p += -P.K*( (P.r_n*l_n+P.r_w*l_w)*P.g )
  bulk_p   =  P.q_w+P.q_n
  dBulk_s  = 0.5*P.K*l_n*dp_c*grad(s_n)
  dBulk_s += P.K*l_n*(grad(p_avg)-P.r_n*P.g)
  bulk_s   = P.q_n

  #### dg penalty factors
  lambdaMax = k('+')*k('-')/avg(k)
  p_c0bis,dp_c0bis,l_n0,l_w0 = P.p_c(0.5)
  penalty_p = [beta*lambdaMax*sMax(l_n0+l_w0), beta*k*(l_n0+l_w0)]
  penalty_s = [0.5*beta*lambdaMax*sMax(l_n0*dp_c0bis), 0.5*beta*k*(l_n0*dp_c0bis)]
\end{python}

\subsection{P-adaptivity}
\label{sec:codeModifications_padaptive}

To change the marking strategy for the p-adaptivity the function
\pyth{markp} needs to be redefined:

\begin{python}
  def markp(element):
      polorder = spc.localOrder(element)
      if element.level < maxLevel: return min(polorder+1,maxOrder)
      val = pEstimator(element,element.referenceElement.center)
      val = [estimate.localFunction(e).evaluate(x)[0],
             estimate_pm1.localFunction(e).evaluate(x)[0]]
      if val[0] > val[1]:
          return polorder-1 if polorder > 1 else polorder
      elif val[0] < 0.01*val[1]:
          return polorder+1 if polorder < maxOrder else polorder
      return polorder
\end{python}

\subsection{Isotropic Flow over weak Lens}
\label{sec:codeModifications_iwl}
Changing the set up of the problem requires modifying the static components
of the problem class i.e. for the isotropic setting with the weaker lens we
need to change permeability tensors:
\begin{python}
  Kdiag = Lens.lens(1e-12, 1e-10)      # [m^2]
  Koff  = Lens.lens(0,0)               # [m^2]
  K     = as_matrix( [[Kdiag,Koff],[Koff,Kdiag]] )
\end{python}

\end{document}